\documentclass[pra,twocolumn,showkeys,showpacs,amsmath,amsfonts]{revtex4-1}
\usepackage[latin2]{inputenc}
\usepackage{graphicx}
\def\be{\begin{equation}}
\def\ee{\end{equation}}
\def\bea{\begin{eqnarray}}
\def\eea{\end{eqnarray}}
\def\bi{\begin{itemize}}
\def\ei{\end{itemize}}

\begin{document}

\title{    Variational tensor network renormalization in imaginary time:\\
           Two-dimensional quantum compass model at finite temperature
}

\author{Piotr Czarnik}
\affiliation{Instytut Fizyki im. Mariana Smoluchowskiego, Uniwersytet Jagiello\'nski,
             prof. S. {\L}ojasiewicza 11, PL-30-348 Krak\'ow, Poland}

\author{Jacek Dziarmaga}
\affiliation{Instytut Fizyki im. Mariana Smoluchowskiego, Uniwersytet Jagiello\'nski,
             prof. S. {\L}ojasiewicza 11, PL-30-348 Krak\'ow, Poland}

\author{Andrzej M. Ole\'{s} }
\affiliation{Max-Planck-Institut f\"ur Festk\"orperforschung,
             Heisenbergstrasse 1, D-70569 Stuttgart, Germany}
\affiliation{Instytut Fizyki im. Mariana Smoluchowskiego, Uniwersytet Jagiello\'nski,
             prof. S. {\L}ojasiewicza 11, PL-30-348 Krak\'ow, Poland}

\date{\today}

\begin{abstract}
Progress in describing thermodynamic phase transitions in quantum
systems is obtained by noticing that the Gibbs operator $e^{-\beta H}$
for a two-dimensional (2D) lattice system with a Hamiltonian $H$ can be
represented by a three-dimensional tensor network, the third dimension
being the imaginary time (inverse temperature) $\beta$. Coarse-graining
the network along $\beta$ results in a 2D projected entangled-pair
operator (PEPO) with a finite bond dimension $D$. The coarse-graining
is performed by a tree tensor network of isometries. The isometries are
optimized variationally --- taking into account full tensor environment
--- to maximize the accuracy of the PEPO. The algorithm is applied to
the isotropic quantum compass model on an infinite square lattice near 
a symmetry-breaking phase transition at finite temperature. From the 
linear susceptibility in the symmetric phase and the order parameter 
in the symmetry-broken phase the critical temperature is estimated at 
${\cal T}_c=0.0606(4)J$, where $J$ is the isotropic coupling constant 
between $S=1/2$ pseudospins.
\end{abstract}


\maketitle

%%%%%%%%%%%%%%%%%%%%%%%%%%%%%%%%%%%%%%%%%%%%%%%%%%%%%%%%%%%%%%%%%%%%%%%%%%
\section{Introduction}
%%%%%%%%%%%%%%%%%%%%%%%%%%%%%%%%%%%%%%%%%%%%%%%%%%%%%%%%%%%%%%%%%%%%%%%%%%

Understanding phase transitions and broken symmetries in frustrated
many-body quantum systems remains one of the major challenges of
modern physics. Frustration in magnetic systems occurs by competing
exchange interactions and leads frequently to disordered spin liquids
\cite{Nor09,Bal10}. However, this does not happen in two-dimensional
(2D) classical systems where ordered states with broken $\mathbb{Z}_2$
symmetry occur at finite temperature, as in the exactly solvable Ising
models with fully frustrated lattice, or with frustration distributed
periodically along columns \cite{Vil77,Lon80}.
In contrast, quantum spins interacting by SU(2) symmetric interactions
order only at zero temperature in the 2D Heisenberg model. Whether or
not 2D quantum spin models with interactions of lower symmetry do
order at finite temperature is a challenging problem in the theory.
Unfortunately, quantum spin systems interacting on a square lattice
are not exactly solvable as entanglement plays an important role
\cite{Ami08}, and advanced methods which deal with entangled
degrees of freedom have to be applied.

Perhaps the simplest example of frustrated quantum exchange
interactions is found in the 2D compass model \cite{Nus15}, where two
different spin components interact along horizontal or vertical bonds
of the square lattice. Recent interest in the compass models is
motivated by spin-orbital physics in transition metal oxides with
active orbital degrees of freedom
\cite{Kug82,Fei97,Ole05,Kha05,Ulr06,Kru09,Karlo,Woh11,Brz12,Brz15}.
This field is very challenging due to the interplay and entanglement
of spins and orbitals which leads to remarkable consequences in real
materials \cite{Ole12}. However, when spin order is ferromagnetic or
when spin and orbitals couple strongly by spin-orbit interaction
\cite{Jac09}, the exchange interactions simplify and concern only
orbitals or pseudopins. A generic model which stands for all these
situations is the 2D compass model. It represents directional
orbital interactions between $e_g$ or $t_{2g}$ orbitals on the bonds
in a 2D square or three-dimensional (3D) cubic lattice
\cite{vdB99,vdB04,Ish05,Ryn10,Wro10,Dag10,Tro13,Che13}.
Its better understanding is crucial not only for spin-orbital systems
but also for its realizations in optical lattices \cite{Mil07}.
Unlike the spins interacting by Heisenberg SU(2) symmetric exchange,
the 2D compass model for orbitals breaks the symmetry at finite
temperature in form of nematic order \cite{Wen08}. It is remarkable
that in nanoscopic systems this order survives perturbing Heisenberg
interactions in the lowest energy excited states, providing a
perspective for its applications in quantum computing \cite{Tro10}.
A better understanding of the signatures of this phase transition
provides a theoretical challenge.

To address these questions we develop below tensor network
renormalization at finite temperature, following the pioneering work
by two of us \cite{var}. The quantum tensor networks proved to be a
competitive tool to study strongly correlated quantum systems
\cite{Che15}. Their advent was a discovery of the density matrix
renormalization group (DMRG) \cite{White,Sch05} that was later shown to
optimize the matrix product state (MPS) variational \textit{Ansatz}
\cite{Sch11}. Over the last decade, MPS was generalized to a 2D
projected entangled pair state (PEPS) \cite{PEPS} and supplemented with
the multiscale entanglement renormalization \textit{Ansatz} (MERA)
\cite{MERA}. As variational methods these networks do not suffer from
the fermionic sign problem \cite{fermions} and fermionic PEPS provided
the most accurate results for the $t-J$ \cite{PEPStJ} and Hubbard
\cite{CorbozHubbard} models employed to study the high-${\cal T}_c$
superconductivity. The networks --- both MPS
\cite{WhiteKagome,CincioVidal,starDMRG} and PEPS
\cite{PepsRVB,PepsKagome,PepsJ1J2} --- made also some major
breakthroughs in the search for topological order. This is where
geometric frustration often prohibits the traditional quantum Monte
Carlo (QMC).

Thermal states of quantum Hamiltonians were explored much less than
their ground states. In one-dimensional (1D) models they can be
represented by MPS \textit{Ansatz}
prepared by accurate imaginary time evolution \cite{ancillas,WhiteT}.
A similar approach can be applied in 2D case \cite{Czarnik,self} --- 
the PEPS manifold is a compact representation for Gibbs states 
\cite{Molnar} --- but the accurate evolution proved to be more 
challenging there. Alternative direct contractions of the 3D partition 
function were proposed \cite{ChinaT} but, due to local tensor update, 
they are expected to converge more slowly with increasing refinement 
parameter. Even a small improvement towards the full update can 
accelerate the convergence significantly \cite{HOSRG}. This
research parallels similar progress in finite temperature variational
Monte Carlo, see e.g. Ref. \cite{japMC}.

In order to overcome these problems, two of us introduced a variational
algorithm to optimize a finite temperature projected entangled-pair
operator (PEPO) \cite{var}. The 3D network $e^{-\beta H}$ is
coarse-grained along the imaginary time $\beta$ (inverse temperature)
to obtain the PEPO \textit{Ansatz} for $e^{-\beta H}$. The 
coarse-graining is optimized variationally --- employing full/nonlocal 
tensor environments --- in order to maximize the accuracy of the 
coarse-grained PEPO. A benchmark application to the 2D quantum Ising 
model in transverse field was presented in Ref. \cite{var}. In this 
paper we move near the edge of geometric frustration and apply the 
same algorithm to the 2D isotropic quantum compass model \cite{Nus15}.
Our results supplement earlier QMC studies \cite{Tanaka,Wen08,Wen10},
and a high-temperature expansion \cite{Oit11} studies concerning the
symmetry breaking phase transition in this model which happens at 
finite temperature.

This paper is organized as follows. In Sec. \ref{sec:compass} we
introduce the 2D quantum compass model and summarize the results on
its finite temperature symmetry-breaking phase transition. In Sec.
\ref{sec:algorithm} the algorithm for variational renormalization is
described in detail, but some more technical features are delegated to
Appendices \ref{sec:CMR}, \ref{sec:Error}, and \ref{sec:Px}. They
include the standard corner matrix renormalization in Appendix
\ref{sec:CMR} as well as new elements, like a direct estimate of the
error inflicted by the finite bond dimension in Appendix \ref{sec:Error}
and variational optimization in case of non-symmetric environments
introduced in Appendix \ref{sec:Px}. The numerical results obtained for
the 2D quantum compass model are collected in Sec. \ref{sec:results}.
We analyze the order parameter and the susceptibility in Sec.
\ref{sec:op} as well as spin-spin correlations in Sec. \ref{sec:ss}.
Concluding remarks and a short summary are presented in Sec.
\ref{sec:conclusion}.

%%%%%%%%%%%%%%%%%%%%%%%%%%%%%%%%%%%%%%%%%%%%%%%%%%%%%%%%%%%%%%%%%%%%%%%%%%%%%%%%%%%%%%%%%%%
\section{Quantum compass model}
\label{sec:compass}
%%%%%%%%%%%%%%%%%%%%%%%%%%%%%%%%%%%%%%%%%%%%%%%%%%%%%%%%%%%%%%%%%%%%%%%%%%%%%%%%%%%%%%%%%%%

The quantum compass model on an infinite square lattice
\cite{Nus15} is
\begin{equation}
H = -\frac14 J_x \sum_j X_{j}X_{j+e_a} -\frac14 J_z \sum_j Z_{j}Z_{j+e_b}.
\label{H}
\end{equation}
Here $j$ is a site number and $X_j\equiv\sigma^x_j$ and 
$Z_j\equiv\sigma^z_j$ are Pauli matrices at site $j$, and $e_a(e_b)$ 
are unit vectors along the $a(b)$ axis. The model is a sum of nearest 
neighbor Ising-type ferromagnetic couplings between $S=1/2$ 
pseudospins: $J_xX_jX_{j+e_a}/4$ for a bond along the $a$ axis and 
$J_zZ_jZ_{j+e_b}/4$ along the $b$ axis. We consider mainly the 
isotropic case, and set $J_x=J_z=J=1$. The order parameter is,
\be
Q \equiv
\left|\left\langle Q_j \right\rangle\right|=
\left|\left\langle X_{j}X_{j+e_a} - Z_{j}Z_{j+e_b}\right\rangle\right|.
\label{OP}
\ee
For convenience we define $Q\geq 0$, i.e., for the cases when $Q_j<0$
we transform the obtained state to $Q_j>0$ by exchanging simultaneously
the two axes and the two spin components, $a\leftrightarrow b$ and 
$X\leftrightarrow Z$. The order parameter is finite below the phase 
transition that occurs at temperature ${\cal T}_c$. This transition 
belongs to the $d=2$ Ising universality class \cite{Wen10,Mis04}.

Recent progress in understanding the nature of nematic order in the
2D quantum compass model is due to the uncovering the consequences of
its symmetries. It was shown that the spectral properties can be
uniquely determined by discrete symmetries like parity \cite{Brz13}.
The conservation of spin parities in rows and columns in the 2D
quantum compass model (for $x$ and $z$-components of spins) has very
interesting consequences. While the most of the two-site spin
correlations vanish in the ground state, the two-dimer correlations
exhibit the nontrivial hidden order \cite{Brz13}.

The phase transition to such an exotic nematic state with hidden order
was studied with QMC, and its critical temperature was estimated at
${\cal T}_c=0.0585$ \cite{Wen10}. As compared to the classical
compass model, it is strongly suppressed by quantum fluctuations
\cite{Wen08}.
A high-temperature series expansion in $\beta$ up to order $\beta^{24}$
predicted \cite{Oit11} --- using an extrapolation with Pad\'e
approximants --- a similar but (estimated to be) less accurate value
${\cal T}_c=0.0625$. The same extrapolation, but with the ${\cal T}_c$
fixed at the QMC value, estimated the susceptibility exponent
$\gamma\simeq 1.3$ that is close to the exact $\gamma=1.75$ but slightly
away from it. In this paper we readdress these questions with the tensor
network algorithm presented below.

%%%%%%%%%%%%%%%%%%%%%%%%%%%%%%%%%%%%%%%%%%%%%%%%%%%%%%%%%%%%%%%%%%%%%%%%%%
\section{ALGORITHM}
\label{sec:algorithm}
%%%%%%%%%%%%%%%%%%%%%%%%%%%%%%%%%%%%%%%%%%%%%%%%%%%%%%%%%%%%%%%%%%%%%%%%%%

In this Section we describe the algorithm that was introduced and
tested for the 2D quantum Ising model in Ref. \cite{var}. Here we
present its less symmetric version suitable for the compass model.
Unlike in the Ising model, where results could be easily converged by
increasing a PEPO bond dimension $D$, here they require an extrapolation
with $1/D\to0$. In Appendix \ref{sec:Error} we explain how to estimate
the error inflicted by a finite $D$. The extrapolation becomes
smoother when $1/D$ is replaced by the error estimate.

%%%%%%%%%%%%%%%%%%%%%%%%%%%%%%%%%%%%%%%%%%%%%%%%%%%%%%%%%%%%%%%%%%%%%%%%%%
\subsection{Purification of thermal states}
\label{sec:purification}
%%%%%%%%%%%%%%%%%%%%%%%%%%%%%%%%%%%%%%%%%%%%%%%%%%%%%%%%%%%%%%%%%%%%%%%%%%

We consider spins-$1/2$ with a Hamiltonian $H$ on an infinite square
lattice. Every spin has states numbered by an index $s=0,1$ and is
accompanied by an ancilla with states $a=0,1$. The enlarged
``spin+ancilla'' space is spanned by states $\prod_j |s_j,a_j\rangle$,
where $j$ is the index of a lattice site. The Gibbs operator at an
inverse temperature $\beta$ is obtained from its purification
$|\psi(\beta)\rangle$ in the enlarged space by tracing out the ancillas,
\be
\rho(\beta) \propto
e^{-\beta H} =
{\rm Tr}_{\rm ancillas}|\psi(\beta)\rangle\langle\psi(\beta)|.
\label{rhobeta}
\ee
At $\beta=0$ we choose a product over lattice sites,
\be
|\psi(0)\rangle = \prod_j ~\sum_{s=0,1} |s_j,s_j\rangle ,
\label{psi0}
\ee
to initialize the imaginary time evolution,
\be
|\psi(\beta)\rangle~=~
e^{-\frac12\beta H}|\psi(0)\rangle\equiv
U(\beta)|\psi(0)\rangle.
\label{psibeta}
\ee
The gate $U(\beta)=e^{-\frac12\beta H}$ acts in the Hilbert space of spins.
With the initial state (\ref{psi0}) the trace in Eq. (\ref{rhobeta}) yields
\be
\rho(\beta) ~\propto~ U(\beta)U^\dag(\beta).
\label{UU}
\ee
$U(\beta)$ will be represented by a PEPO.

%%%%%%%%%%%%%%%%%%%%%%%%%%%%%%%%%%%%%%%%%%%%%%%%%%%%%%%%%%%%%%%%%%%%%%%%%%
\subsection{Suzuki-Trotter decomposition}
\label{sec:ST}
%%%%%%%%%%%%%%%%%%%%%%%%%%%%%%%%%%%%%%%%%%%%%%%%%%%%%%%%%%%%%%%%%%%%%%%%%%
We define gates
\bea
U_{XX}(d\beta) &\equiv &
\prod_{\langle j,j'\rangle||a}e^{\frac{d\beta}{8}X_jX_{j'}},
\nonumber~\\
U_{ZZ}(d\beta) &\equiv &
\prod_{\langle j,j'\rangle||b}e^{\frac{d\beta}{8}Z_jZ_{j'}}.
\label{UXXUZZ}
\eea
In the second-order Suzuki-Trotter decomposition an infinitesimal gate
can be approximated in two ways:
\bea
U(d\beta/2)&\approx &U_{XX}(d\beta/4)U_{ZZ}(d\beta/2)U_{XX}(d\beta/4),
\nonumber\\
U(d\beta/2)&\approx &U_{ZZ}(d\beta/4)U_{XX}(d\beta/2)U_{ZZ}(d\beta/4).
\label{U}
\eea
We combine them into an elementary time step,
\bea
U(d\beta)
&=&
U_{XX}(d\beta/4)
U_{ZZ}(d\beta/2)
U_{XX}(d\beta/4)\nonumber\\
&\times&
U_{ZZ}(d\beta/4)
U_{XX}(d\beta/2)
U_{ZZ}(d\beta/4).
\label{Udbeta}
\eea

%%%%%%%%%%%%%%%%%%%%%%%%%%%%%%%%%%%%%%%%%%%%%%%%%%%%%%%%%%%%%%%%%%%%%%%%%%%%
\begin{figure}[t!]
\vspace{-0cm}
\includegraphics[width=0.9\columnwidth,clip=true]{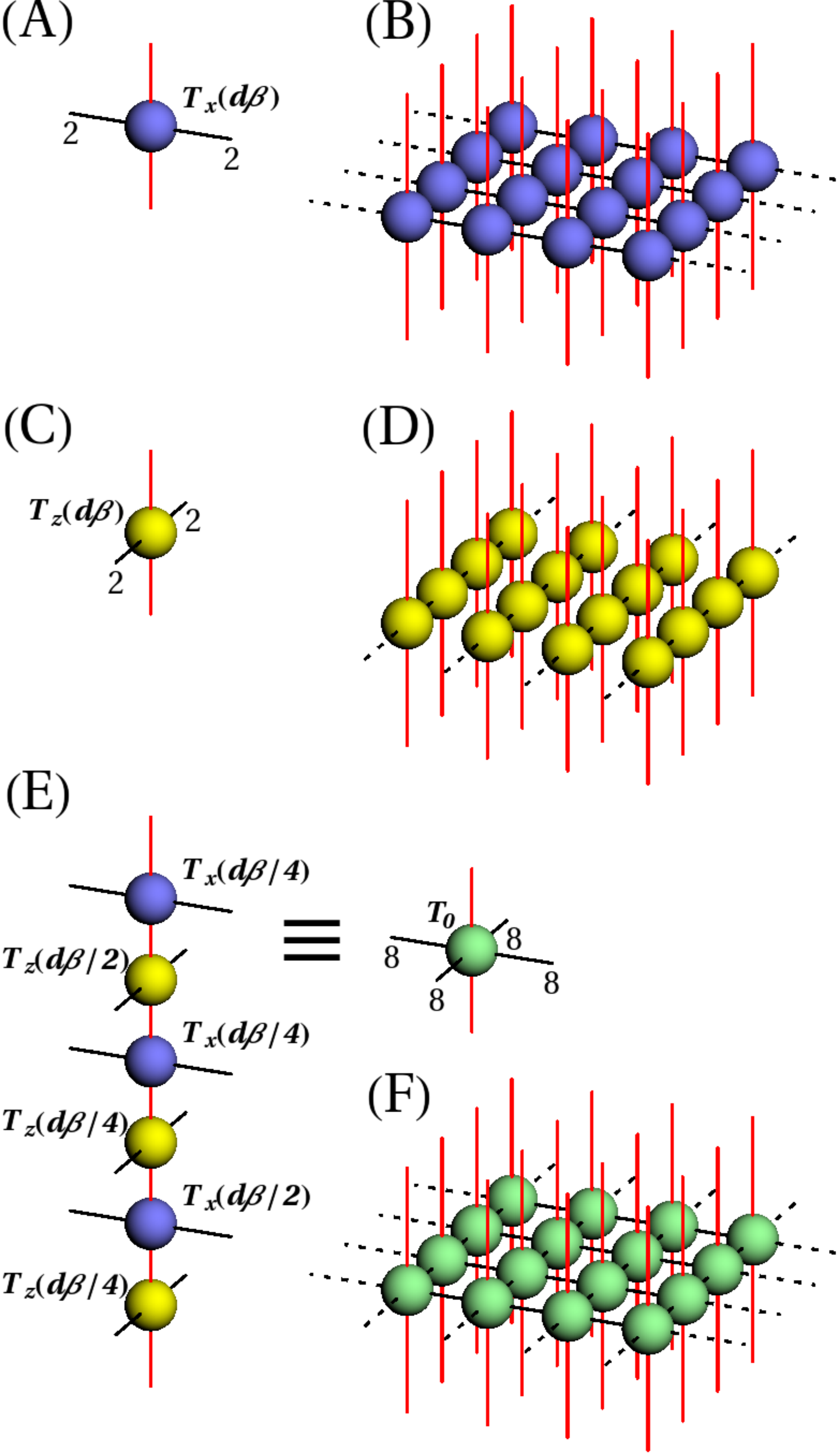}
\vspace{-0cm}
\caption{
In A,
the Trotter tensor $T_x(d\beta)$ with two spin indices (red lines)
and two bond indices (black lines) along the $a$ axis.
The bond indices have (bond) dimension $2$.
In B,
the gate $U_{XX}(d\beta)$ is a layer of tensors $T_x(d\beta)$
contracted through their bond indices.
In C,
$T_z(d\beta)$ with bond indices along the $b$ axis.
In D,
the gate $U_{ZZ}(d\beta)$ is a layer of $T_z(d\beta)$ contracted
through their bond indices.
In E,
the six Trotter tensors contributing to the elementary time step
$U(d\beta)$ in Eq. (\ref{Udbeta}) can be merged into a single
elementary Trotter tensor $T_0$ with a bond dimension $8$.
In F,
the time step $U(d\beta)$ is a layer of tensors $T_0$.
}
\label{fig:TxTzT0}
\end{figure}

To rearrange $U(d\beta)$ as a tensor network, at every bond in
Eqs. (\ref{UXXUZZ}) we make a singular value decomposition
\bea
e^{\frac{d\beta}{8}X_jX_{j'}} &=&
\sum_{\mu=0,1}
x_{j,\mu}~
x_{j',\mu}, \nonumber\\
e^{\frac{d\beta}{8}Z_jZ_{j'}} &=&
\sum_{\mu=0,1}
z_{j,\mu}~
z_{j',\mu}.
\eea
Here $\mu$ is a bond index,
$x_{j,\mu}\equiv\sqrt{\Lambda_\mu}\,(X_j)^\mu$,
and
$z_{j,\mu}\equiv\sqrt{\Lambda_\mu}\,(Z_j)^\mu$.
The singular values are $\Lambda_0=\cosh\frac{d\beta}{8}$ and
$\Lambda_1=\sinh\frac{d\beta}{8}$. Now we can write
\bea
U_{XX}(d\beta)
&=&
\sum_{\{\mu\}}
\prod_j
\left(
\prod_{j'}
x_{j,\mu_{\langle j,j'\rangle}}
\right).
\label{Tx}
\eea
Here $\mu_{\langle j,j'\rangle}$ is a bond index for the NN bond
$\langle j,j'\rangle$ along $a$ axis, and $\{\mu\}$ is a set of all such
bond indices. The brackets enclose a Trotter tensor $T_x(d\beta)$ at
site $j$, see Fig. \ref{fig:TxTzT0}A. It is a spin operator depending
on bond indices connecting its site with its two NNs along the $a$ axis.
A contraction of these Trotter tensors is the gate $U_{XX}(d\beta)$ in
Fig. \ref{fig:TxTzT0}B. In a similar way,
\bea
U_{ZZ}(d\beta)
&=&
\sum_{\{\mu\}}
\prod_j
\left(
\prod_{j'}
z_{j,\mu_{\langle j,j'\rangle}}
\right).
\label{Tz}
\eea
Here the brackets enclose a Trotter tensor $T_z(d\beta)$ at site $j$,
shown in Fig. \ref{fig:TxTzT0}C.
A layer of these Trotter tensors is the gate $U_{ZZ}(d\beta)$ in Fig.
\ref{fig:TxTzT0}D.

To represent the time step (\ref{Udbeta}) in Fig. \ref{fig:TxTzT0}E,
six Trotter tensors are contracted along imaginary time into an
elementary Trotter tensor $T_0$.
Along each bond there are $3$ bond indices of dimension $2$ that are
combined into a single one of dimension $8$. A layer of $T_0$ in Fig.
\ref{fig:TxTzT0}F is the time step (\ref{Udbeta}).

The evolution operator is a product of $N$ such elementary time steps,
\be
U(\beta)=\left[U(d\beta)\right]^N,
\label{UN}
\ee
where $N=\beta/d\beta$ is a number of time steps. So far the only
approximation is the Suzuki-Trotter decomposition.

%%%%%%%%%%%%%%%%%%%%%%%%%%%%%%%%%%%%%%%%%%%%%%%%%%%%%%%%%%%%%%%%%%%%%%%%%%
\subsection{Coarse graining and renormalization in imaginary time}
\label{sec:CG}
%%%%%%%%%%%%%%%%%%%%%%%%%%%%%%%%%%%%%%%%%%%%%%%%%%%%%%%%%%%%%%%%%%%%%%%%%%

\begin{figure}[t!]
\vspace{-0cm}
\includegraphics[width=0.9\columnwidth,clip=true]{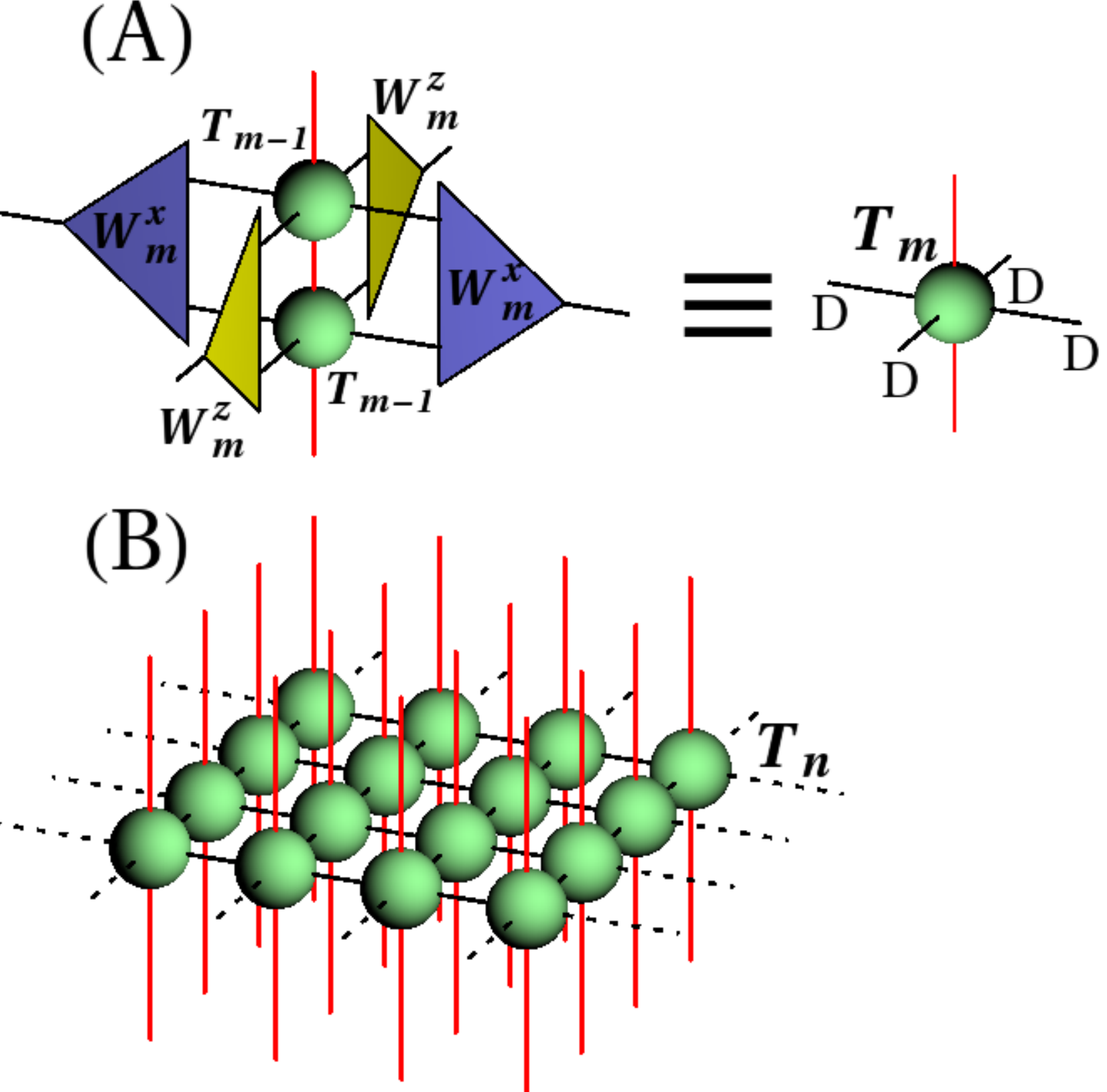}
\vspace{-0cm}
\caption{
In A,
the coarse-graining step along the imaginary time.
Two Trotter tensors $T_{m-1}$ are combined and then renormalized into
a single tensor $T_m$. The renormalization is made by isometries
$W_m^x$ and $W_m^z$ on the bonds along the $a$ axis and $b$ axis,
respectively. In B, after $n$ coarse-graining tranformations the PEPO
tensor $T_n$ is obtained. A layer of contracted $T_n$ makes the PEPO
\textit{Ansatz} for the gate $U(\beta)$. It is equivalent to the PEPS
\textit{Ansatz} for the purification $|\psi(\beta)\rangle$
when its bottom spin indices are reinterpreted as ancilla indices.
}
\label{fig:Tm}
\end{figure}

\begin{figure}[b!]
\vspace{-0cm}
\includegraphics[width=0.99\columnwidth,clip=true]{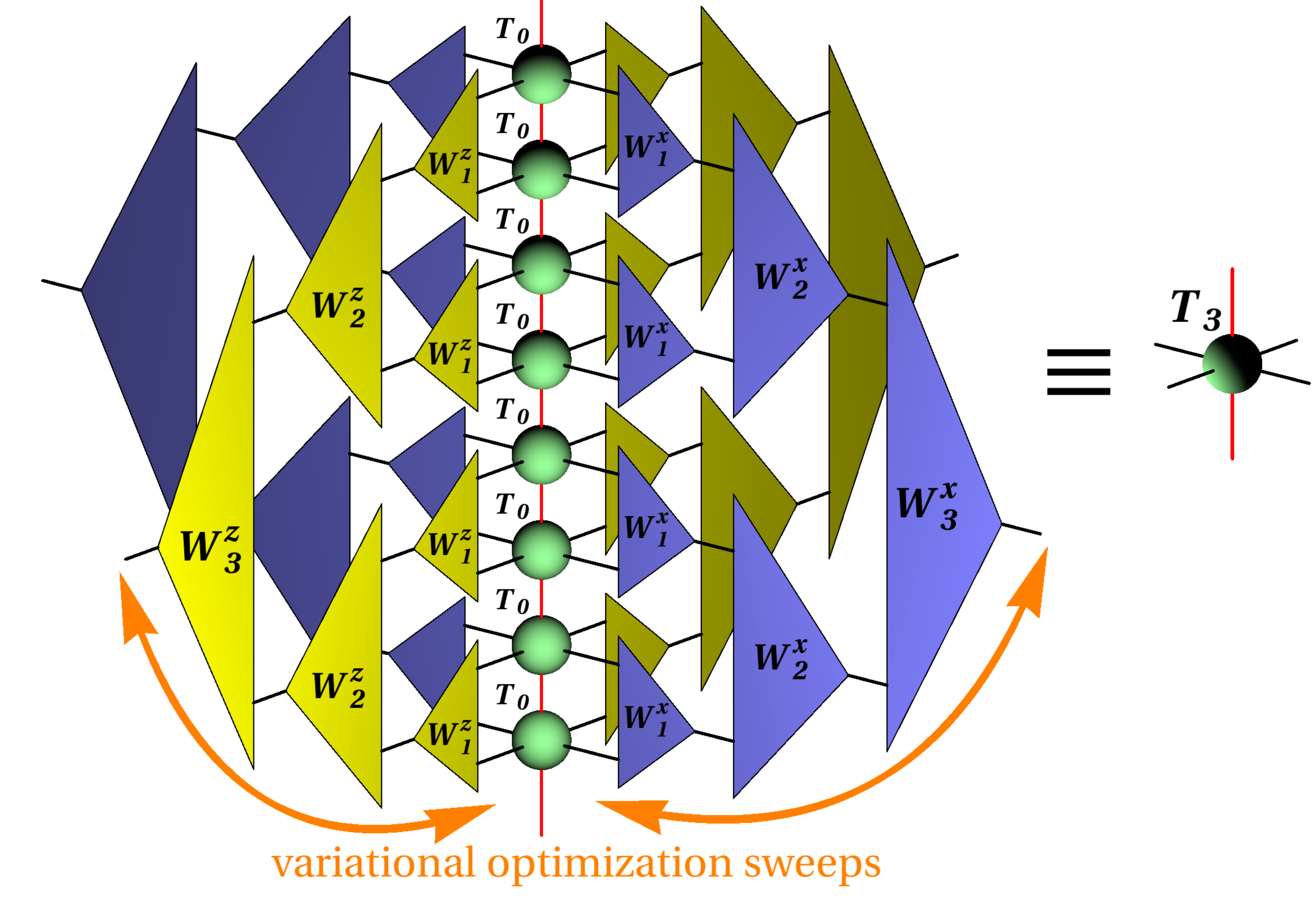}
\vspace{-0cm}
\caption{
Three coarse-graining transformations result in the Trotter tensor $T_3$.
The isometries acting along a given bond combine into a tree tensor
network (TTN). They are optimized by repeated up- and down-sweeps.
}
\label{fig:TTN}
\end{figure}

Equation (\ref{UN}) suggests to combine $N$ elementary tensors $T_0$'s
into a single PEPO tensor in a similar way as in Fig. \ref{fig:TxTzT0}E
the six Trotter tensors were combined into a single $T_0$.
Unfortunately,
along each bond this would require to combine
$N$ bond indices of dimension $8$ into a single one of dimension $8^N$.
To prevent this exponential blow-up with $N$,
we proceed step by step each time combining just two tensors into one:
$T_0\times T_0\to T_1$,...,$T_{n-1}\times T_{n-1}\to T_n$.
Here
\be
n=\log_2N=\log_2 \frac{\beta}{d\beta}
\ee
is the total number of the coarse-graining transformations that is only
logarithmic in the total number of Suzuki-Trotter steps $N$
(and logarithmic in the small time step $d\beta$). After each step,
$T_{m-1}\times T_{m-1}\to T_m$, the combined bond indices are
renormalized down to $D$ by isometries $W_m$.
The indices along the $a$ axis are renormalized by isometries $W^x_m$
and those along the $b$ axis by $W^z_m$, see Fig. \ref{fig:Tm}A.
Figure \ref{fig:TTN} shows the net outcome after $m=3$ coarse-graining
transformations. Along each bond there are $3$ layers of isometries,
from $W_1$ to $W_3$,
that combine into a tree tensor network (TTN) \cite{TTN}.

%%%%%%%%%%%%%%%%%%%%%%%%%%%%%%%%%%%%%%%%%%%%%%%%%%%%%%%%%%%%%%%%%%%%%%%%%%%%

%%%%%%%%%%%%%%%%%%%%%%%%%%%%%%%%%%%%%%%%%%%%%%%%%%%%%%%%%
%%%%%%%%%%%%%%%%%%%%%%%%%%%%%%%%%%%%%%%%%%%%%%%%%%%%%%%%%%%%%%%%%%%%%%%%%%%%

%%%%%%%%%%%%%%%%%%%%%%%%%%%%%%%%%%%%%%%%%%%%%%%%%%%%%%%%%%%%%%%%%%%%%%%%%%%
\begin{figure}[t!]
\vspace{-0cm}
\includegraphics[width=0.9\columnwidth,clip=true]{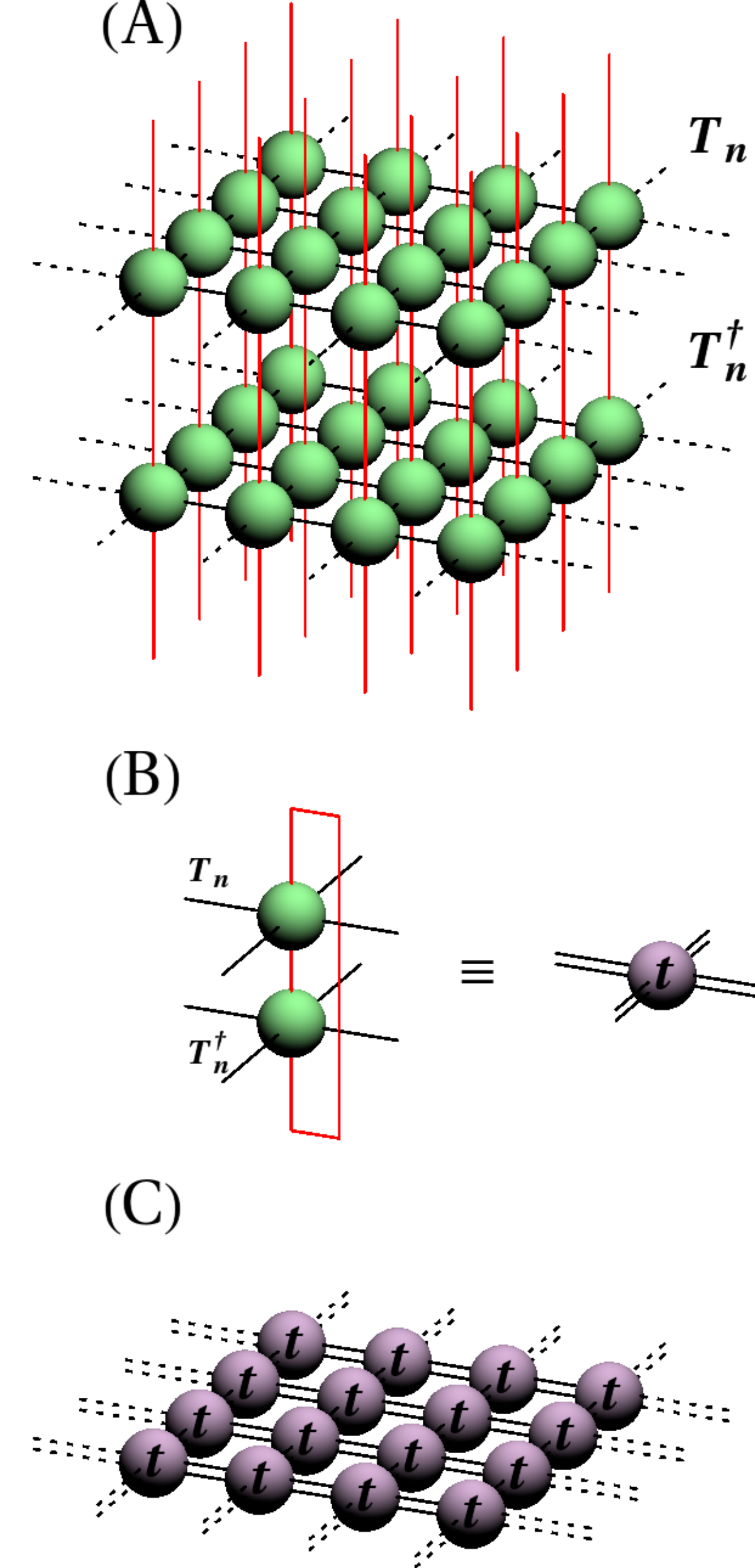}
\vspace{-0cm}
\caption{
In A,
the operator $e^{-\beta H}$ obtained after combining the two gates
$U(\beta)$ and $U^\dag(\beta)$
in Fig. \ref{fig:Tm}B according to Eq. (\ref{UU}). Here the layers
of tensors $T_n$ and $T_n^\dag$ represent
$U(\beta)$ and $U^\dag(\beta)$ respectively.
In B,
two tensors $T_n$ combine into a transfer tensor $t$.
In C,
a layer of contracted transfer tensors is the partition function
$Z={\rm Tr}e^{-\beta H}$.
}
\label{fig:Tn}
\end{figure}
%%%%%%%%%%%%%%%%%%%%%%%%%%%%%%%%%%%%%%%%%%%%%%%%%%%%%%%%%%%%%%%%%%%%%%%%%%

A layer of $T_n$ shown in Fig. \ref{fig:Tm}B is the PEPO \textit{Ansatz}
for the gate $U(\beta)$. When its bottom spin indices are reinterpreted
as ancilla indices it becomes a PEPS \textit{Ansatz} for the
purification $|\psi(\beta)\rangle$. Figure \ref{fig:Tn}A shows how to
combine two gates $U(\beta)$ into the Gibbs operator $e^{-\beta H}$
according to Eq. (\ref{UU}). A single layer of transfer tensors $t$ in
Figs. \ref{fig:Tn}B and \ref{fig:Tn}C is an \textit{Ansatz} for the
partition function $Z={\rm Tr}~e^{-\beta H}$.

%%%%%%%%%%%%%%%%%%%%%%%%%%%%%%%%%%%%%%%%%%%%%%%%%%%%%%%%%%%%%%%%%%%%%%%%%%%%
\begin{figure}[b!]
\vspace{-0cm}
\includegraphics[width=0.9\columnwidth,clip=true]{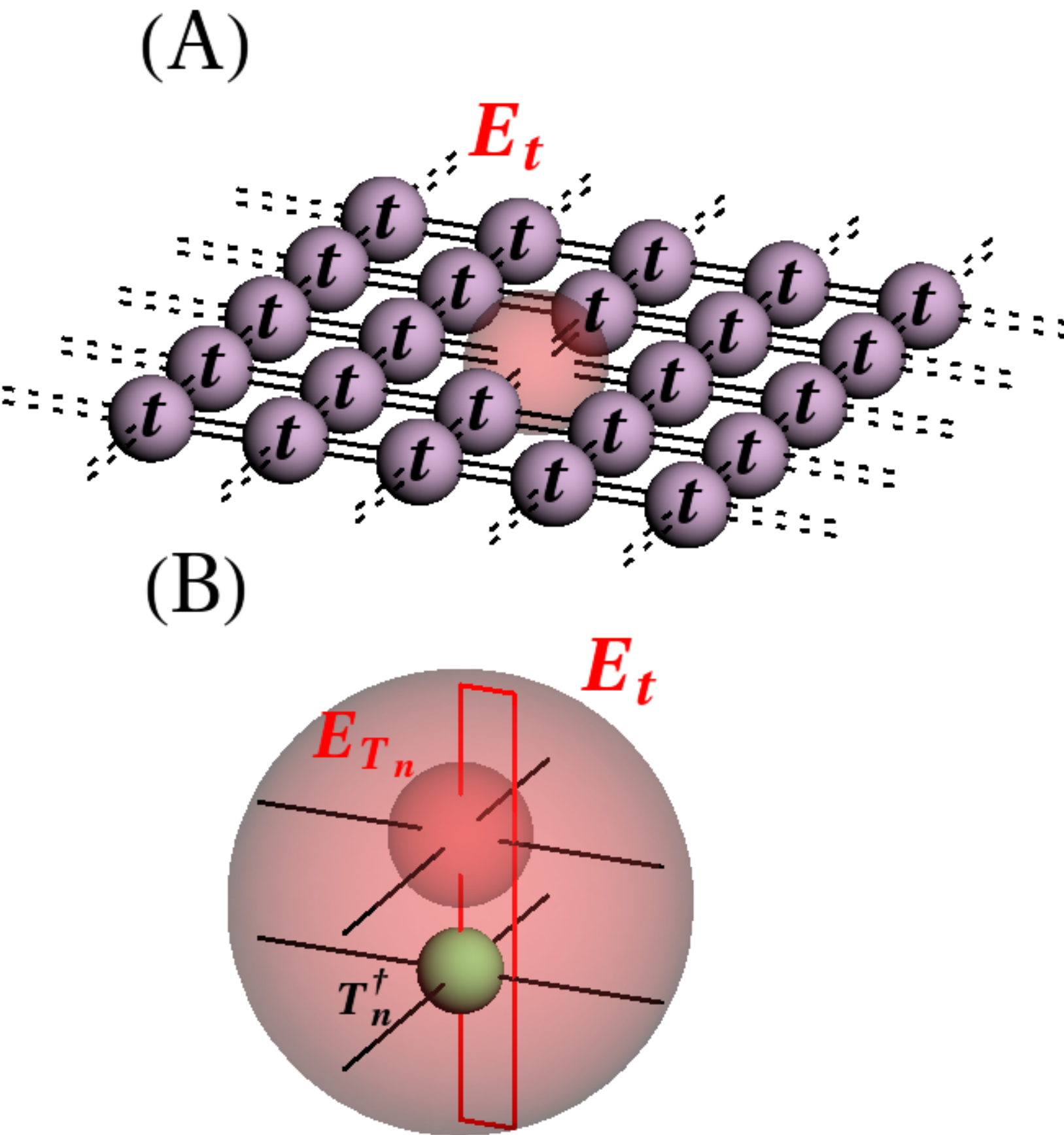}
\vspace{-0cm}
\caption{
In A,
the tensor environment $E_t$ for the tensor $t$ obtained after removing
one tensor $t$ from the partition function in Fig. \ref{fig:Tn}C.
In B,
tensor environment $E_{T_n}$ for the PEPO tensor $T_n$ obtained from $E_t$.
}
\label{fig:Envt}
\end{figure}
%%%%%%%%%%%%%%%%%%%%%%%%%%%%%%%%%%%%%%%%%%%%%%%%%%%%%%%%%%%%%%%%%%%%%%%%%%%%

%%%%%%%%%%%%%%%%%%%%%%%%%%%%%%%%%%%%%%%%%%%%%%%%%%%%%%%%%%%%%%%%%%%%%%%%%%%%
\subsection{Variational optimization}
\label{sec:Var}
%%%%%%%%%%%%%%%%%%%%%%%%%%%%%%%%%%%%%%%%%%%%%%%%%%%%%%%%%%%%%%%%%%%%%%%%%%%%
In order to optimize the isometries we need an efficient algorithm to
calculate a tensor environment of each isometry.
A tensor environment of $W_m$ is the tensor $E_{W_m}$ that is generated
by removing one $W_m$ from the partition function.
It is proportional to the gradient $\partial Z/\partial W_m$.
The algorithm proceeds step by step down the hierarchy of isometries.

%%%%%%%%%%%%%%%%%%%%%%%%%%%%%%%%%%%%%%%%%%%%%%%%%%%%%%%%%%%%%%%%%%%%%%%%%%%%
\begin{figure}[t!]
\vspace{-0cm}
\includegraphics[width=0.9\columnwidth,clip=true]{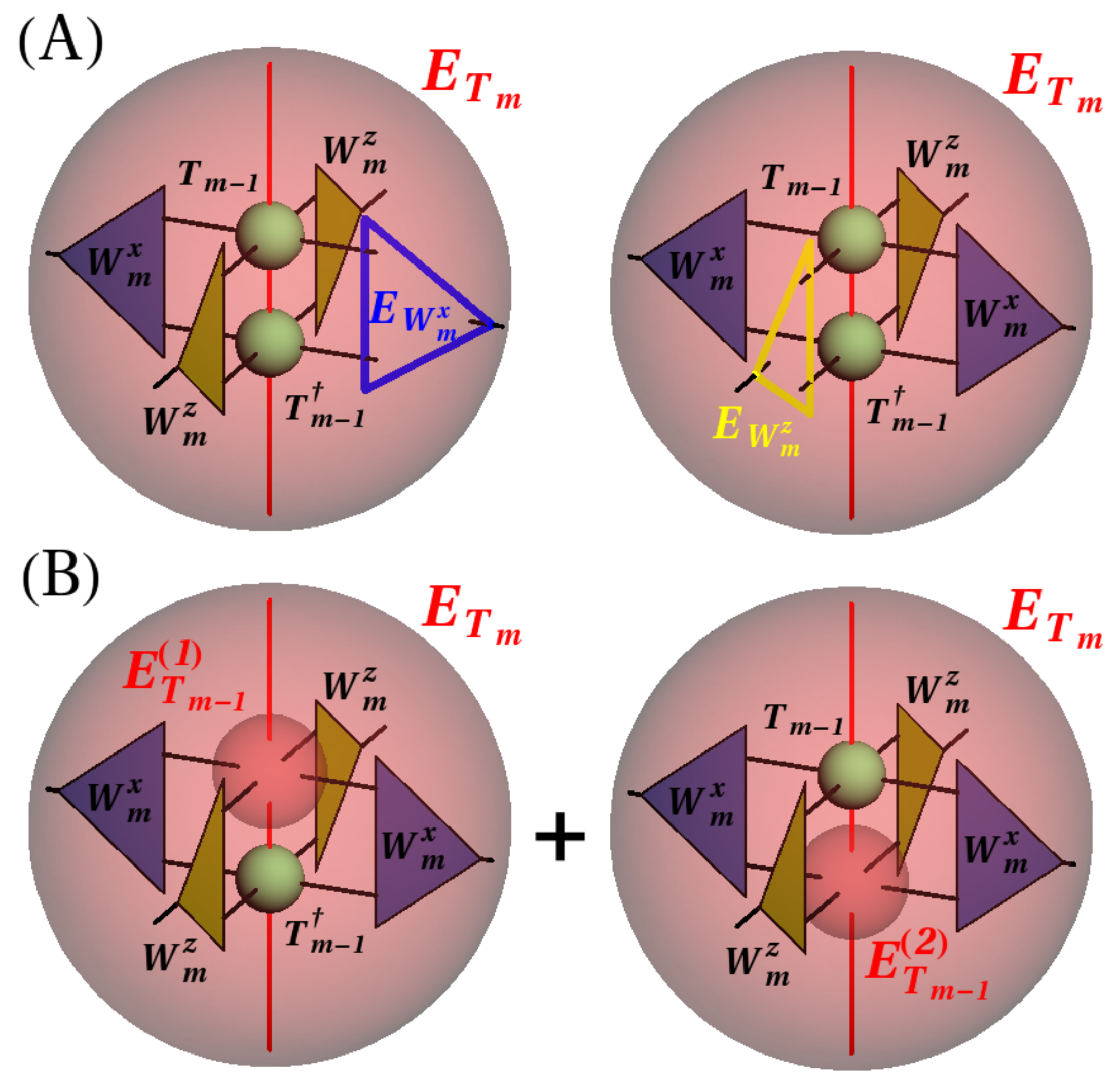}
\vspace{-0cm}
\caption{
In A,
the environments of the isometries $W_m^x$ and $W_m^z$ are obtained
from $E_{T_m}$.
In B,
the step from the environment $E_{T_m}$ down to $E_{T_{m-1}}$.
There are two inequivalent contributions to $E_{T_{m-1}}$.
They add up to $E_{T_{m-1}}=E_{T_{m-1}}^{(1)}+E_{T_{m-1}}^{(2)}$.
}
\label{fig:EnvW}
\end{figure}
%%%%%%%%%%%%%%%%%%%%%%%%%%%%%%%%%%%%%%%%%%%%%%%%%%%%%%%%%%%%%%%%%%%%%%%%%%%%

A preparatory step is calculation of an environment
$E_t\propto\partial Z/\partial t$ of the transfer tensor $t$ in Fig.
\ref{fig:Envt}A. It is the tensor that remains after removing one
transfer tensor from the partition function in Fig. \ref{fig:Tn}C.
The infinite network $E_t$ cannot be contracted exactly,
but its accurate approximation, that can be improved in a systematic
way by increasing a control parameter $M$, can be obtained with the
corner matrix renormalization (CMR) \cite{CMR} described in Appendix
\ref{sec:CMR}.
Once converged, $E_t$ is contracted with one $T_n$
to yield an environment $E_{T_n}$ of the PEPO tensor $T_n$,
see Fig. \ref{fig:Envt}B.
With $E_{T_n}$ we can initialize a down optimization sweep.

\begin{figure}[t!]
\vspace{-0cm}
\includegraphics[width=0.85\columnwidth,clip=true]{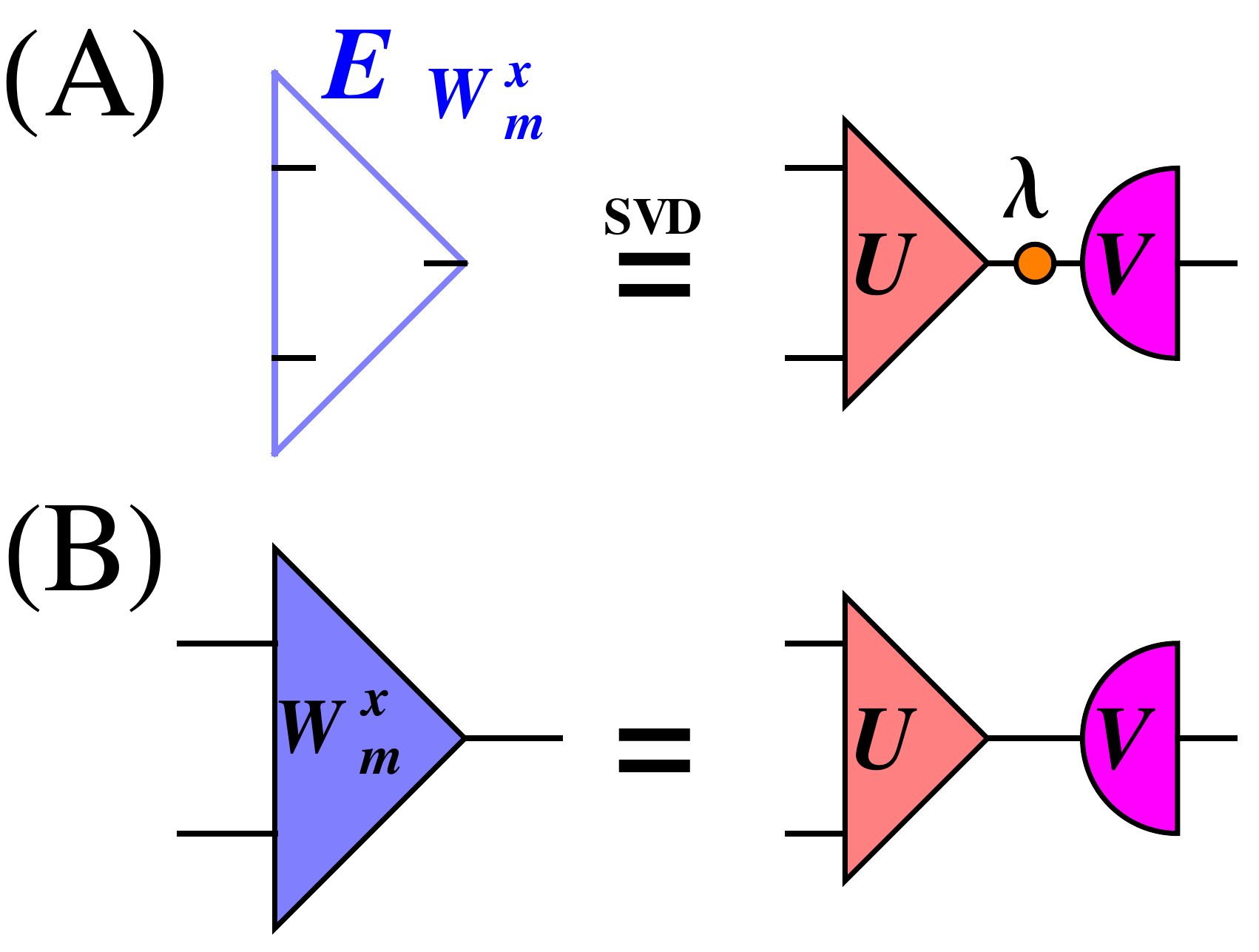}
\vspace{-0cm}
\caption{
The update of the isometry $W^x_m$.
In A,
the isometric environment is subject to a singular value decomposition
(SVD), $E_{W^x_m}=U\lambda V^\dag$.
In B,
the isometry is updated as $W_m^x=UV^\dag$.
A similar procedure is applied to $W^z_m$.
}
\label{fig:svdEnvW}
\end{figure}

From $E_{T_n}$ we obtain the environments $E_{W_n^x}$ and $E_{W_n^z}$,
see Fig. \ref{fig:EnvW}A.
These environments are used immediately to update their isometries,
see Fig. \ref{fig:svdEnvW}.
With the updated $W_n$ we can calculate $E_{T_{n-1}}$,
see Fig. \ref{fig:EnvW}B.
From $E_{T_{n-1}}$ we obtain the environments $E_{W_{n-1}}$ and
use them immediately to update the isometries $W_{n-1}$.
The same procedure is repeated all the way down to $W_1$
whose update completes the down-sweep.

Once $W_1$ were updated, an up optimization sweep begins.
It has $n$ steps.
In the $m$-th step two tensors $T_{m-1}$ and the environment $E_{T_m}$
--- calculated before during the down-sweep --- are contracted to
obtain the environments $E_{W_m}$ and update the isometries $W_m$,
see Fig. \ref{fig:EnvW}B. The updated $W_m$ are used to coarse-grain
$T_{m-1}\times T_{m-1}\to T_m$, see Fig. \ref{fig:Tm}A.
This basic step is repeated all the way up to $T_n$.

The up-sweep completes one optimization loop consisting of three stages:
\bi
\item the CMR procedure:
$$
T_n\,\stackrel{\rm CMR}{\longrightarrow}\, E_t\to\, E_{T_n};
$$
\item the down-sweep:
$$
E_{T_n}\to\, E_{W_n}\to\, E_{T_{n-1}}\to\dots\to\, E_{T_1}\to\, E_{W_1};
$$
\item the up-sweep:
$$
T_0\to\, E_{W_1}\to\, T_1\to\dots\to\, T_{n-1}\to\, E_{W_n}\to\, T_n.
$$
\ei
Here each $E_{W_m}$ is used immediately to update $W_m$. The loop is
repeated until convergence.

%%%%%%%%%%%%%%%%%%%%%%%%%%%%%%%%%%%%%%%%%%%%%%%%%%%%%%%%%%%%%%%%%%%%%%%%%%%%
%%%%%%%%%%%%%%%%%%%%%%%%%%%%%%%%%%%%%%%%%%%%%%%%%%%%%%%%%%%%%%%%%%%%%%%%%%%%

The numerical cost of all the procedures in this Section scales like
$D^8$. Typically, it is sub-leading as compared to the cost of CMR in
Appendix \ref{sec:CMR}. Having thus outlined the algorithm we can
proceed now with the results obtained for the 2D quantum compass model
(\ref{H}).

%%%%%%%%%%%%%%%%%%%%%%%%%%%%%%%%%%%%%%%%%%%%%%%%%%%%%%%%%%%%%%%%%%%%%%%%%%%%%%%
\section{Results}
\label{sec:results}
%%%%%%%%%%%%%%%%%%%%%%%%%%%%%%%%%%%%%%%%%%%%%%%%%%%%%%%%%%%%%%%%%%%%%%%%%%%%%%%

%%%%%%%%%%%%%%%%%%%%%%%%%%%%%%%%%%%%%%%%%%%%%%%%%%%%%%%%%%%%%%%%%%%%%%%%%%%%%%%
\subsection{Order parameter and its susceptibility}
\label{sec:op}
%%%%%%%%%%%%%%%%%%%%%%%%%%%%%%%%%%%%%%%%%%%%%%%%%%%%%%%%%%%%%%%%%%%%%%%%%%%%%%%

The mean extrapolated values of the order parameter in the
symmetry-broken phase were fitted with the scaling function
\be
Q({\cal T})\propto({\cal T}_c-{\cal T})^{\beta},
\ee
where $\beta$ stands here for the critical exponent of the order
parameter (this notation is widely accepted and we use it in this
Section only). In this way the critical temperature was estimated
as ${\cal T}_c=0.06090$, where the number of digits indicates
precision of the linear fit alone. The exponent was estimated here
as $\beta=0.223$ that is close but somewhat removed from the exact
$\beta=1/8$.

In the symmetric phase on the other
side of the transition, the mean extrapolated values of the linear
susceptibility were fitted with
\be
\chi({\cal T})\propto({\cal T}-{\cal T}_c)^\gamma,
\label{chi}
\ee
where $\gamma$ is the susceptibility exponent.
The susceptibility is defined as
\be
\chi=\left.\frac{dQ}{dA}\right|_{A=0},
\label{chiA0}
\ee
where $A$ is the anisotropy of the coupling constants in Eq. (\ref{H}):
$J_x=1+A/2$ and $J_z=1-A/2$.
The derivative (\ref{chiA0}) was approximated accurately by a finite
difference between $A=10^{-5}$ and $A=0$.
The fit (\ref{chi}) yields ${\cal T}_c=0.06021$ and $\gamma=1.35$.
The exponent is again somewhat removed from the exact $\gamma=1.75$.
The estimate of ${\cal T}_c$ is close to that obtained from the order
parameter on the other side of the critical point.

%%%%%%%%%%%%%%%%%%%%%%%%%%%%%%%%%%%%%%%%%%%%%%%%%%%%%%%%%%%%%%%%%%%%%%%%%%%%%%%%%%
\begin{figure}[t!]
%\centering
%\centering
\begin{center}
\includegraphics[width=\linewidth]{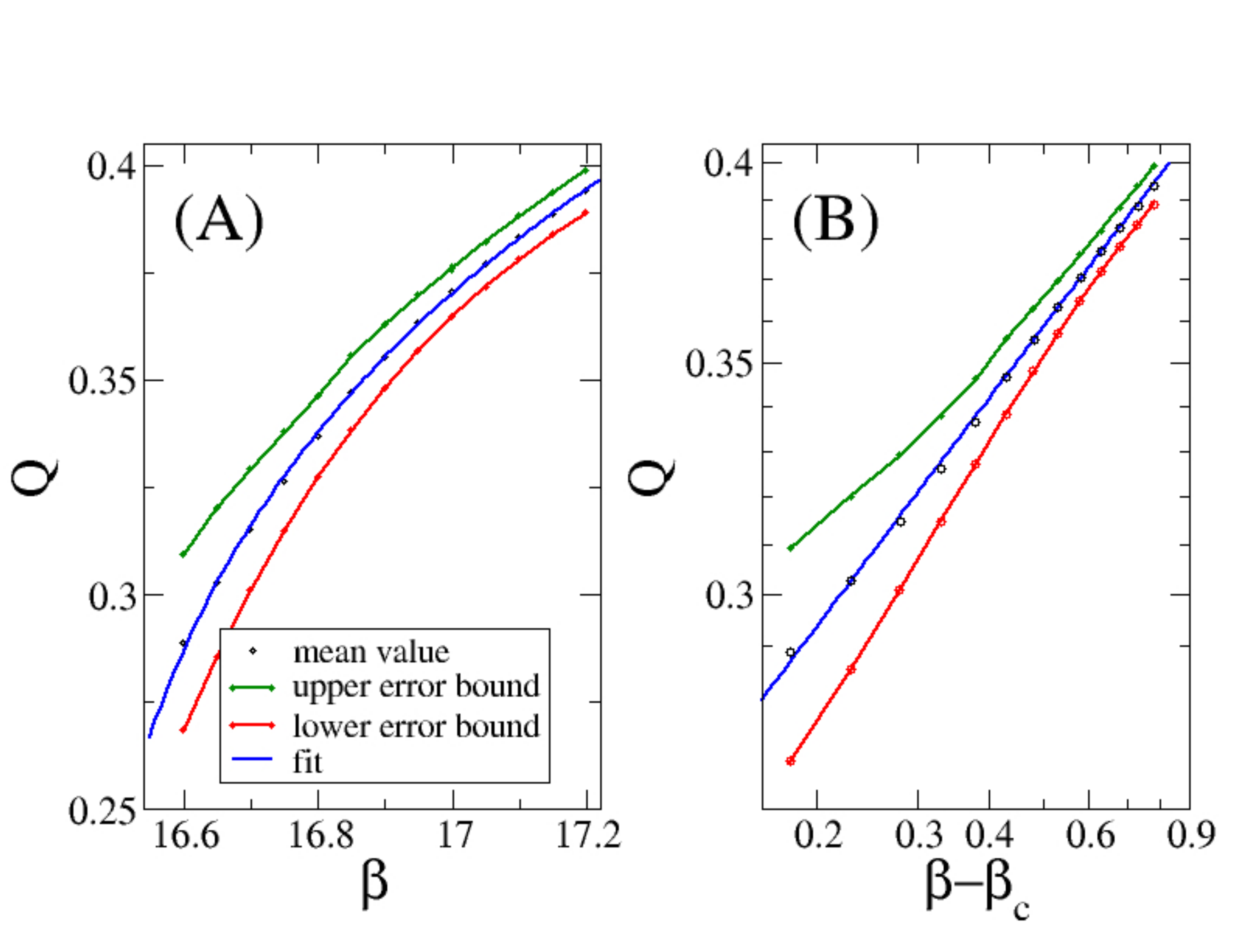}
\end{center}
\caption{
In (A),
the order parameter $Q$ in Eq. (\ref{OP}) in the symmetry-broken phase
near the phase transition. The mean value was obtained by extrapolation
of the renormalization error to $0$, see Appendix \ref{sec:Error} and
Fig. \ref{fig:ordfits}. The error bounds show the errors of the
extrapolation.
The mean value was fitted with $Q\propto({\cal T}_c-{\cal T})^\beta$,
where $\beta=0.223$ is the order parameter exponent
and the critical temperature ${\cal T}_c=0.06090$.
In (B),
a log-log plot of the mean value and the best fit.
}
\label{fig:ord}
\end{figure}
%%%%%%%%%%%%%%%%%%%%%%%%%%%%%%%%%%%%%%%%%%%%%%%%%%%%%%%%%%%%%%%%%%%%%%%%%%%%%%%%%%

Relatively large errors of the critical exponents $\beta$ and $\gamma$
originate from estimates made relatively far from the critical point.
Due to the non-analiticity at the critical point, even a tiny error in
the estimate of ${\cal T}_c$ translates into a large error of a critical
exponent.

\begin{figure}[t!]
\centering
\centering
\includegraphics[width=\linewidth]{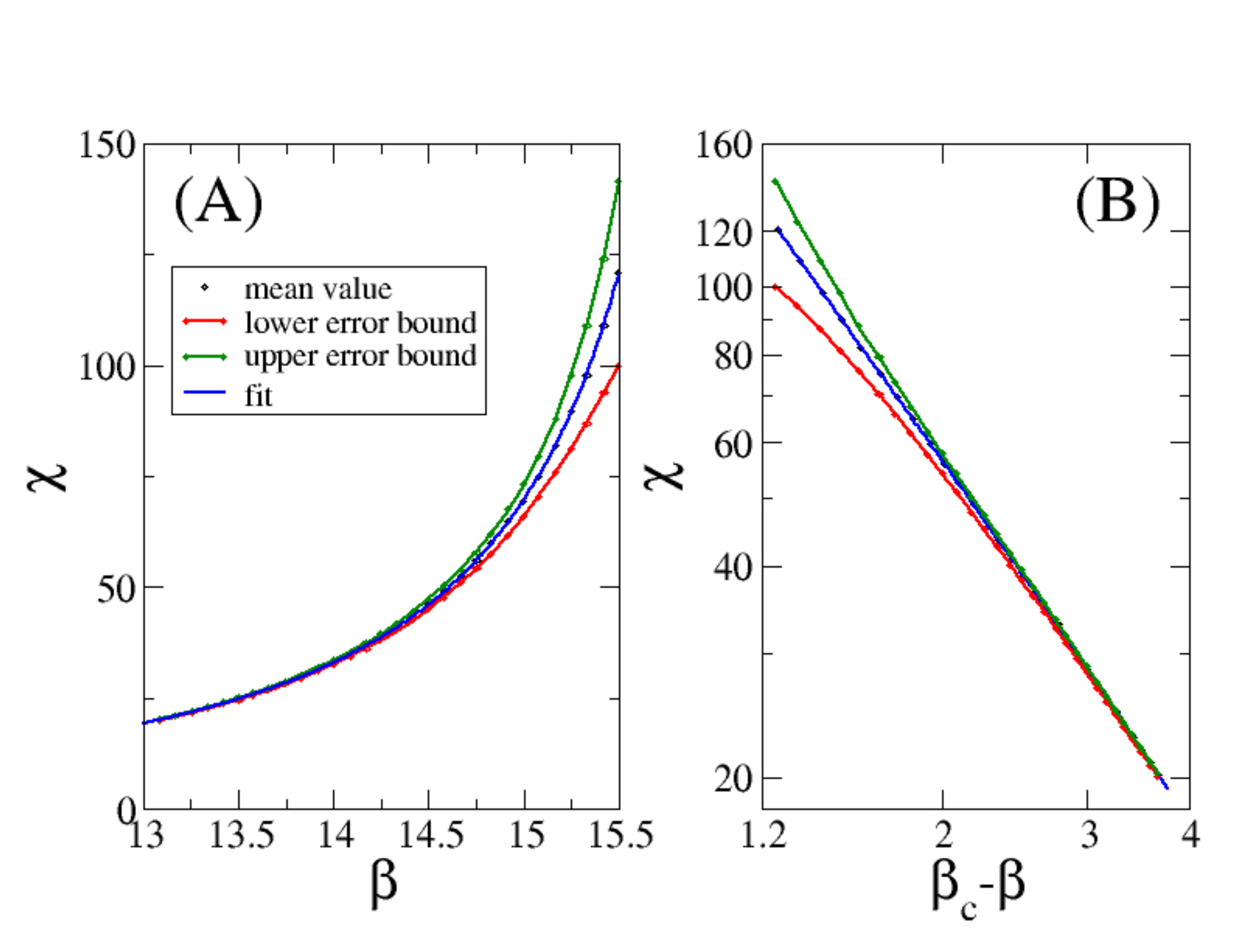}
\caption{
In (A),
the linear susceptibility $\chi$ in \eqref{chi} of the order parameter in
the symmetric phase. The mean value was obtained by extrapolation of
the renormalization error to $0$, see Appendix \ref{sec:Error} and Fig.
\ref{fig:susfits}. The error bounds show the errors of the
extrapolation.
The mean was fitted with $\propto({\cal T}-{\cal T}_c)^\gamma$,
where $\gamma=1.35$ is the susceptibility exponent and
${\cal T}_c=0.06021$.
In (B),
a log-log plot of the mean value and the best fit.
}
\label{fig:sus}
\end{figure}

\begin{figure}[b!]
\centering
\centering
\includegraphics[width=9.2cm]{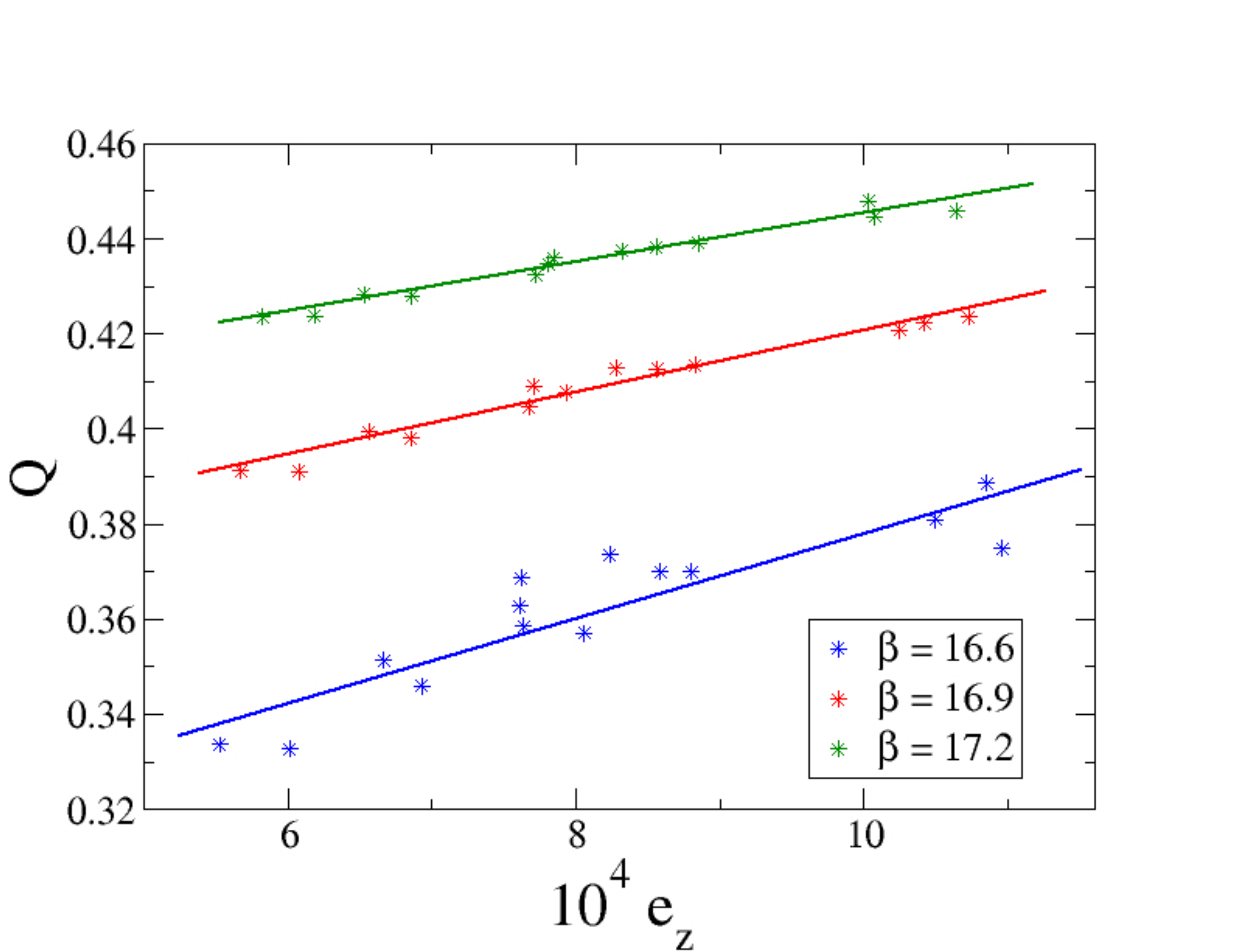}
\caption{
Three examples of the extrapolation of the order parameter $Q$
\eqref{OP} described in Appendix \ref{sec:Error}, to the vanishing
renormalization error, $e_z\to0$.
Here the decreasing $e_z$ corresponds to increasing $D=8,\dots,15$.
The results are converged for the regime of $M\leq40$.
The quality of the linear fit decreases with $\beta$ decreasing towards
the phase transition. Each fit is used to extrapolate to $e_z=0$ and
estimate the errors of the extrapolation.
The means and error bounds are shown in Fig. \ref{fig:ord}.
}
\label{fig:ordfits}
\end{figure}

\begin{figure}[t!]
\centering
\centering
\includegraphics[width=\columnwidth]{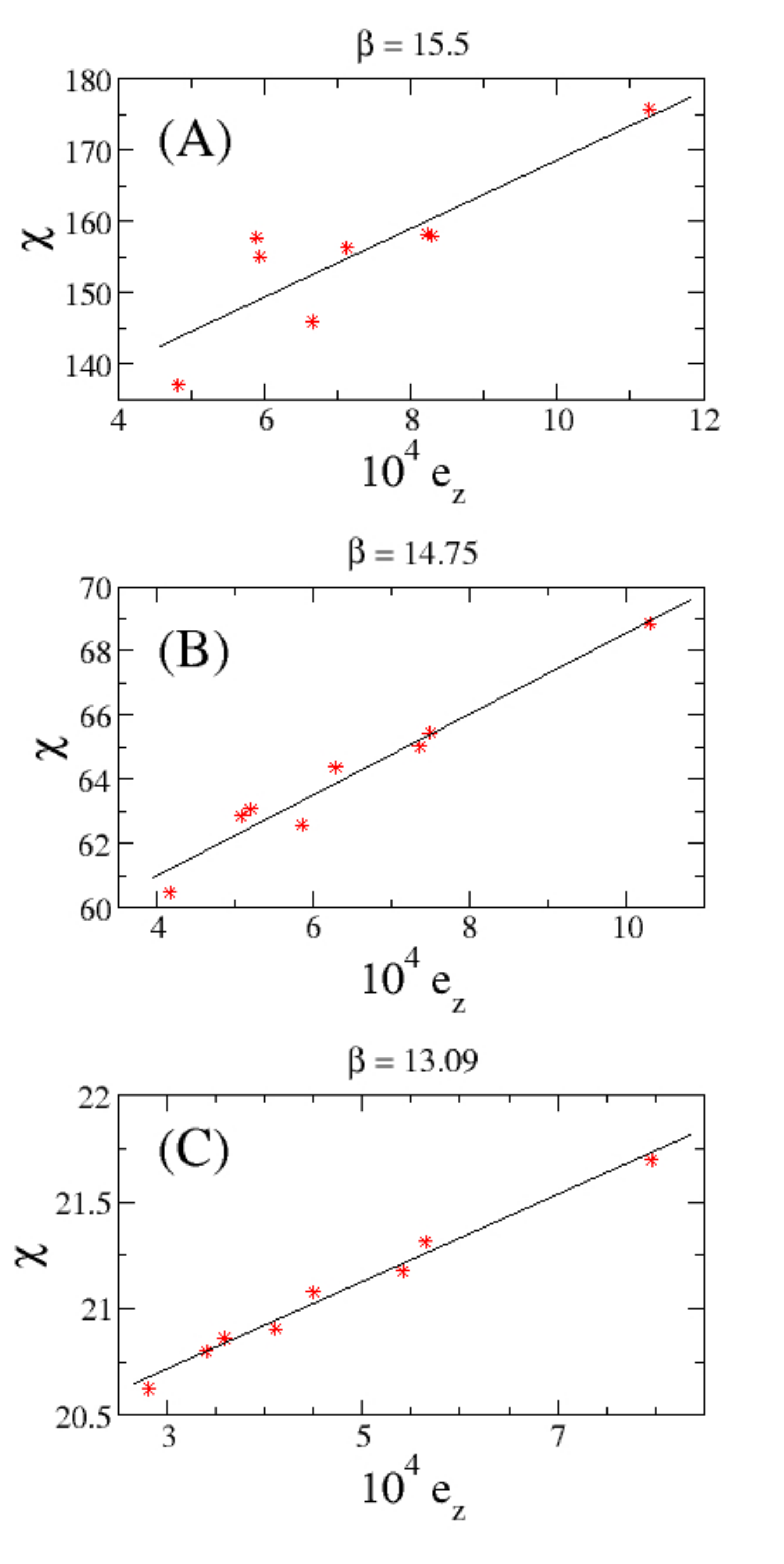}
\caption{
Three examples of the extrapolation of the susceptibility,
described in Appendix \ref{sec:Error},
to the vanishing renormalization error, $e_z\to0$, for:
(A) $\beta=15.5$,
(B) $\beta=14.75$, and
(C) $\beta=13.09$.
Here the decreasing $e_z$ corresponds to increasing $D=8,\dots,15$.
The results are converged for the regime of $M\leq40$. The quality
of the linear fit deteriorates with $\beta$ increasing towards the
phase transition. Each fit is used to extrapolate to $e_z=0$ and to
estimate the errors of this extrapolation, shown together with the
mean value in Fig.~\ref{fig:sus}.
}
\label{fig:susfits}
\end{figure}

Figures \ref{fig:ord} and \ref{fig:sus} show the order parameter and its
linear susceptibility as a function of inverse temperature $\beta$ in
the symmetry-broken and symmetric phases, respectively. The results are
converged in the environmental bond dimension for $M\leq40$, but they
are not quite converged in the bond dimension $D\leq15$. As explained in
Appendix \ref{sec:Error}, instead of the straightforward extrapolation
with $1/D\to0$, it is more reliable to make a smoother extrapolation
with the actual error inflicted by the finite $D$. What is more,
we found that an extrapolation with only the dominant error $e_z\to 0$
is smoother [we recall that $Q>0$ by our convention \eqref{OP}].
We expect that, at least away from the critical point, physical
quantities are analytical in $e_z$ and, consequently, for small enough
$e_z$ they become linear. This expectation is confirmed by our data.
Examples of linear fits used for the extrapolation are shown in
Figs. \ref{fig:ordfits} and \ref{fig:susfits}.
These fits include data for $D=8,\dots,15$. For some $D$ there are more
than one data points corresponding to different random tensors used
to initialize the variational optimization.
Since $e_z$ does not capture all relevant errors --- for instance,
it does not control the accuracy of the environmental tensors ---
it is not justified to keep only the smallest $e_z$ for each $D$.
The quality of the linear fits
decreases when the critical point is approached from either side.

%%%%%%%%%%%%%%%%%%%%%%%%%%%%%%%%%%%%%%%%%%%%%%%%%%%%%%%%%%%%%%%%%%%%%%%%%%%%%%
%%%%%%%%%%%%%%%%%%%%%%%%%%%%%%%%%%%%%%%%%%%%%%%%%%%%%%%%%%%%%%%%%%%%%%%%%%%%%%

%%%%%%%%%%%%%%%%%%%%%%%%%%%%%%%%%%%%%%%%%%%%%%%%%%%%%%%%%%%%%%%%%%%%%%%%%%%%%%%%

%%%%%%%%%%%%%%%%%%%%%%%%%%%%%%%%%%%%%%%%%%%%%%%%%%%%%%%%%%%%%%%%%%%%%%%%%%%%%
%%%%%%%%%%%%%%%%%%%%%%%%%%%%%%%%%%%%%%%%%%%%%%%%%%%%%%%%%%%%%%%%%%%%%%%%%%%%%%%%

The two estimates can be combined into a rough confidence interval
${\cal T}_c\in[0.0602,0.0609]$, or equivalently
${\cal T}_c\simeq 0.0606(4)$, giving a better idea of the actual error
of the method than the tiny errors of the linear fits alone. Our result
agrees well with the most reliable quantum Monte Carlo estimate
${\cal T}_c=0.0585(3)$, see Ref. \cite{Wen10}.

%%%%%%%%%%%%%%%%%%%%%%%%%%%%%%%%%%%%%%%%%%%%%%%%%%%%%%%%%%%%%%%%%%%%%%%%%%%%%%%%%
\subsection{Spin-spin correlation functions}
\label{sec:ss}
%%%%%%%%%%%%%%%%%%%%%%%%%%%%%%%%%%%%%%%%%%%%%%%%%%%%%%%%%%%%%%%%%%%%%%%%%%%%%%%%%

In agreement with predictions for any finite temperature \cite{Nus15},
but in contrast with quantum Monte Carlo \cite{Wen10},
we find zero spontaneous magnetization,
\be
\langle X_m \rangle = 0 = \langle Z_m \rangle,
\label{X0Z}
\ee
within the numerical precision of $10^{-5}$. There is neither any local
magnetization nor any long-range order in the spin-spin correlators.

\begin{figure}[t!]
\centering
\centering
\includegraphics[width=\columnwidth]{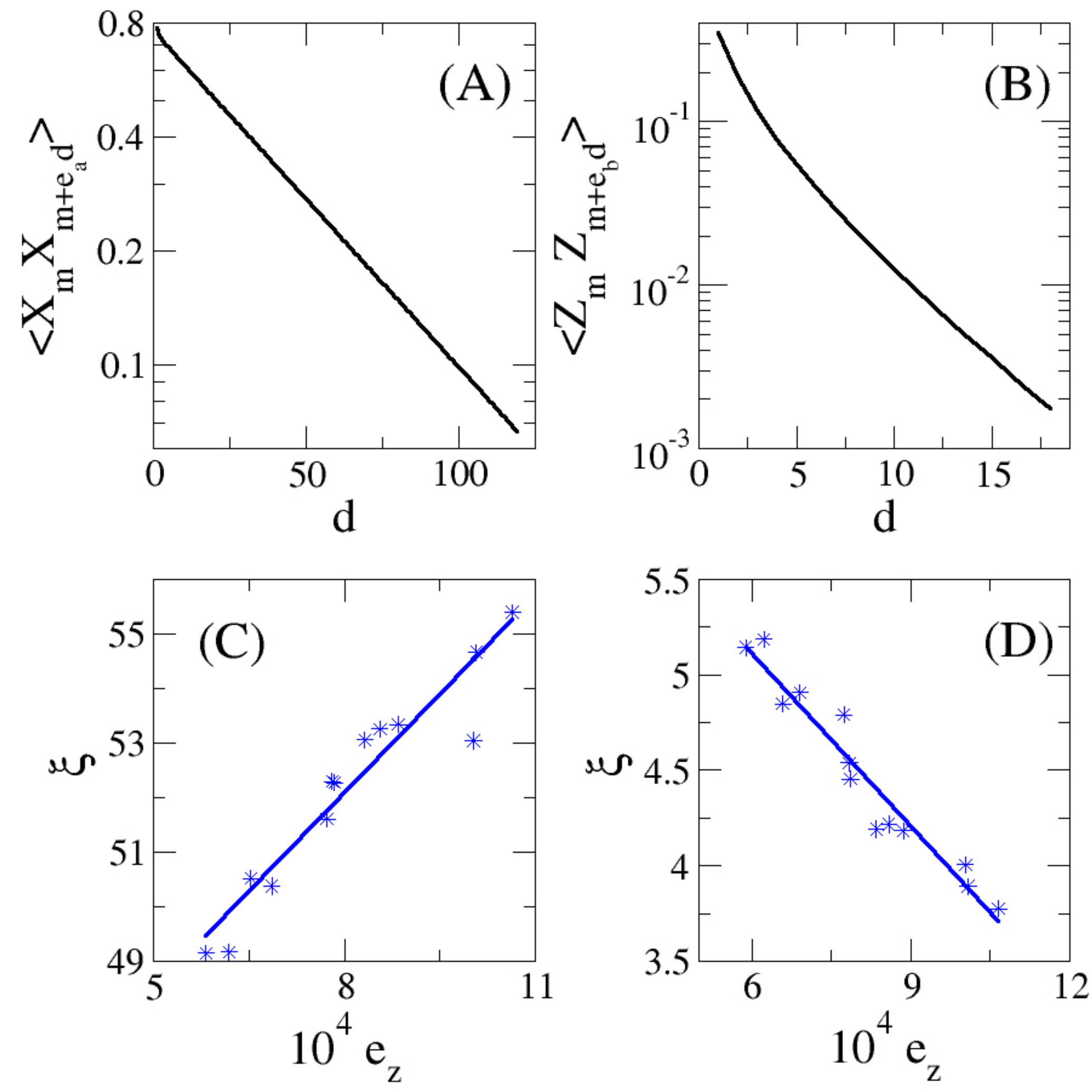}
\caption{
Top --- spin correlations for increasing distance $d$:
(A) the dominant correlation function $\langle X_mX_{m+e_ad}\rangle$ 
along the $a$ axis obtained for $D=15$ and converged in $M$ for $M=40$, 
and
(B) the transverse correlation function $\langle Z_mZ_{m+e_bd}\rangle$ 
along the $b$ axis obtained for $D=15$ and converged in $M$ for $M=60$.
Bottom --- (C)
the dominant correlation length $\xi$ as a function of the error estimate
for different $D$. The extrapolation to zero error gives $\xi=40(2)$, and
(D)
the transverse correlation length $\xi$ as a function of the error.
The extrapolation gives $\xi=6.9(4)$.
}
\label{fig:corr}
\end{figure}

The spin-spin correlations in the symmetry broken phase at $\beta=17.2$
are shown in Fig. \ref{fig:corr}. The dominant correlation function 
along the $a$ axis is exponential but relatively long-ranged with a 
correlation length estimated at $\xi=40(2)$. The transverse correlations 
decay exponentially on a much shorter transverse correlation length 
estimated at $\xi=6.9(4)$.

%%%%%%%%%%%%%%%%%%%%%%%%%%%%%%%%%%%%%%%%%%%%%%%%%%%%%%%%%%%%%%%%%%%%%%%%%%%%%%
%%%%%%%%%%%%%%%%%%%%%%%%%%%%%%%%%%%%%%%%%%%%%%%%%%%%%%%%%%%%%%%%%%%%%%%%%%%%%%%%

\subsection{Numerical details}

All calculations were done in Matlab with an extensive use of the
{\sc ncon} procedure \cite{encon}.
They were checked for convergence in the elementary time step
$d\beta\leq0.005$. The number of isometric layers was fixed at
$n=12$ with the number of time steps $N=2^n=4096$.
To give an idea of the actual time and computer resources needed to
perform the algorithm, the most challenging data points nearest to
the phase transition at $\beta=15.5$ and $\beta=16.6$, with the highest 
bond dimensions $D=15$ and $M=40$, required $1-2$ days on a desktop. 
This time was needed to reach good convergence after $\sim10^2$ 
iterations of the optimization loop. In each loop the CMR procedure 
made tens of iterations to converge the environmental tensors.

At $\beta=15.5$ and $\beta=16.6$, the calculations with $D=15$ were
initialized by embedding converged tensors with smaller $D$ and the
tensors with the smallest $D=8,\dots,10$ were converged after
initialization with random numbers. At each point $100$ simulations
with random initialization were preformed to exclude other solutions.
The calculations farther away from criticality were initialized with
tensors converged closer to it. Additionally, for $D=8,\dots,10$ at
$\beta=15,14.5,14,13.5,13$ and $17.2,17.0,16.8,16.7$ further random
initializations were performed --- $100$ each time --- to exclude 
other solutions. The further away form the phase transition, the fewer
iterations were necessary to reach convergence.

%%%%%%%%%%%%%%%%%%%%%%%%%%%%%%%%%%%%%%%%%%%%%%%%%%%%%%%%%%%%%%%%%%%%%%%%%%%%%%
\section{Conclusion}
\label{sec:conclusion}

\begin{table}[b!]
\centering
\caption{Estimates of the critical temperature ${\cal T}_c$ for the
2D isotropic quantum compass model with $J_x=J_z=1$ as obtained by 
different methods, see Eq. (\ref{H}).}
\begin{ruledtabular}
\begin{tabular}{clccc}
& \hskip.3cm ${\cal T}_c$ &  method  &   Ref.        & \\ \hline
& 0.0625        & high-$T$ expansion & \cite{Oit11}  & \\
& 0.075(2)      & Trotter QMC        & \cite{Tanaka} & \\
& 0.055(1)      & QMC periodic BC    & \cite{Wen08}  & \\
& 0.0585(3)     & QMC screw BC       & \cite{Wen10}  & \\
& 0.0606(4)     & VTNR               &   this work   & \\
\end{tabular}
\end{ruledtabular}
\label{tab:Tc}
\end{table}

We applied the variational tensor network renormalization (VTNR) in
imaginary time, first introduced in Ref. \cite{var}, to the 2D quantum
compass model demonstrating its applicability beyond the quantum Ising
model, in a model of interacting pseudospins close to geometric
frustration.
The method makes efficient use of the bond dimension and it is only
logarithmic in the total number of Suzuki-Trotter imaginary time steps.
An important new algorithmic feature is the extrapolation in the small
error inflicted by the finite bond dimension $D$.

The presented VTNR reproduces the thermodynamic phase transition in
the 2D quantum compass model. In the symmetry broken phase at
${\cal T}<{\cal T}_c$, we find nematic order with long-range spin
correlations along the dominant axis, and short-range correlations in
the transverse direction, but no spontaneous magnetization. We also
attempted to estimate the order parameter exponent $\beta=0.224$ and
the susceptibility exponent $\gamma=1.35$ that are close but somewhat
removed from the exact values $\beta=0.125$ and $\gamma=1.75$,
respectively.

The present approach provides a controlled estimate of the critical
temperature at ${\cal T}_c=0.0606(4)$. In Table \ref{tab:Tc} we
compare this result with earlier estimates including the most recent
QMC \cite{Wen10} with screw boundary conditions (BC). These BC remove
anomalous scalings observed in the case of periodic BC without
introducing the sign problem making Ref. [55] the most reliable
benchmark. Our estimate at $3.5\%$ above their ${\cal T}_c=0.0585(3)$
is in good agreement with QMC. The accuracy of QMC is limited by
extrapolation to infinite system size while the accuracy of our
infinite tensor network by extrapolation to infinite bond dimension.
This positive test suggests that the method used here could be a
competitive tool to treat systems suffering from the sign problem.

%%%%%%%%%%%%%%%%%%%%%%%%%%%%%%%%%%%%%%%%%%%%%%%%%%%%%%%%%%%%%%%%%%%%%%%%%%%%%%
\acknowledgments
%%%%%%%%%%%%%%%%%%%%%%%%%%%%%%%%%%%%%%%%%%%%%%%%%%%%%%%%%%%%%%%%%%%%%%%%%%%%%%
We are indebted to Marek Rams for numerous insightful discussions
and to Anna Francuz for valuable comments on the manuscript.
We kindly acknowledge support by Narodowe Centrum Nauki
(NCN, National Science Center) under Projects:
No. 2013/09/B/ST3/01603 (P.C.~and~J.D.)
and
No. 2012/04/A/ST3/00331 (A.M.O.).
The work of P.C. on his Ph.D. thesis was supported by Narodowe
Centrum Nauki (NCN, National Science Center) under Project
No. 2015/16/T/ST3/00502.

%%%%%%%%%%%%%%%%%%%%%%%%%%%%%%%%%%%%%%%%%%%%%%%%%%%%%%%%%%%%%%%%%%%%%%%%%%%%%%
\appendix
%%%%%%%%%%%%%%%%%%%%%%%%%%%%%%%%%%%%%%%%%%%%%%%%%%%%%%%%%%%%%%%%%%%%%%%%%%%%%%

%%%%%%%%%%%%%%%%%%%%%%%%%%%%%%%%%%%%%%%%%%%%%%%%%%%%%%%%%%%%%%%%%%%%%%%%%%
\section{Corner matrix renormalization}
\label{sec:CMR}
%%%%%%%%%%%%%%%%%%%%%%%%%%%%%%%%%%%%%%%%%%%%%%%%%%%%%%%%%%%%%%%%%%%%%%%%%%

%%%%%%%%%%%%%%%%%%%%%%%%%%%%%%%%%%%%%%%%%%%%%%%%%%%%%%%%%%%%%%%%%%%%%%%%%%
\begin{figure}[t!]
\includegraphics[width=0.99\columnwidth,clip=true]{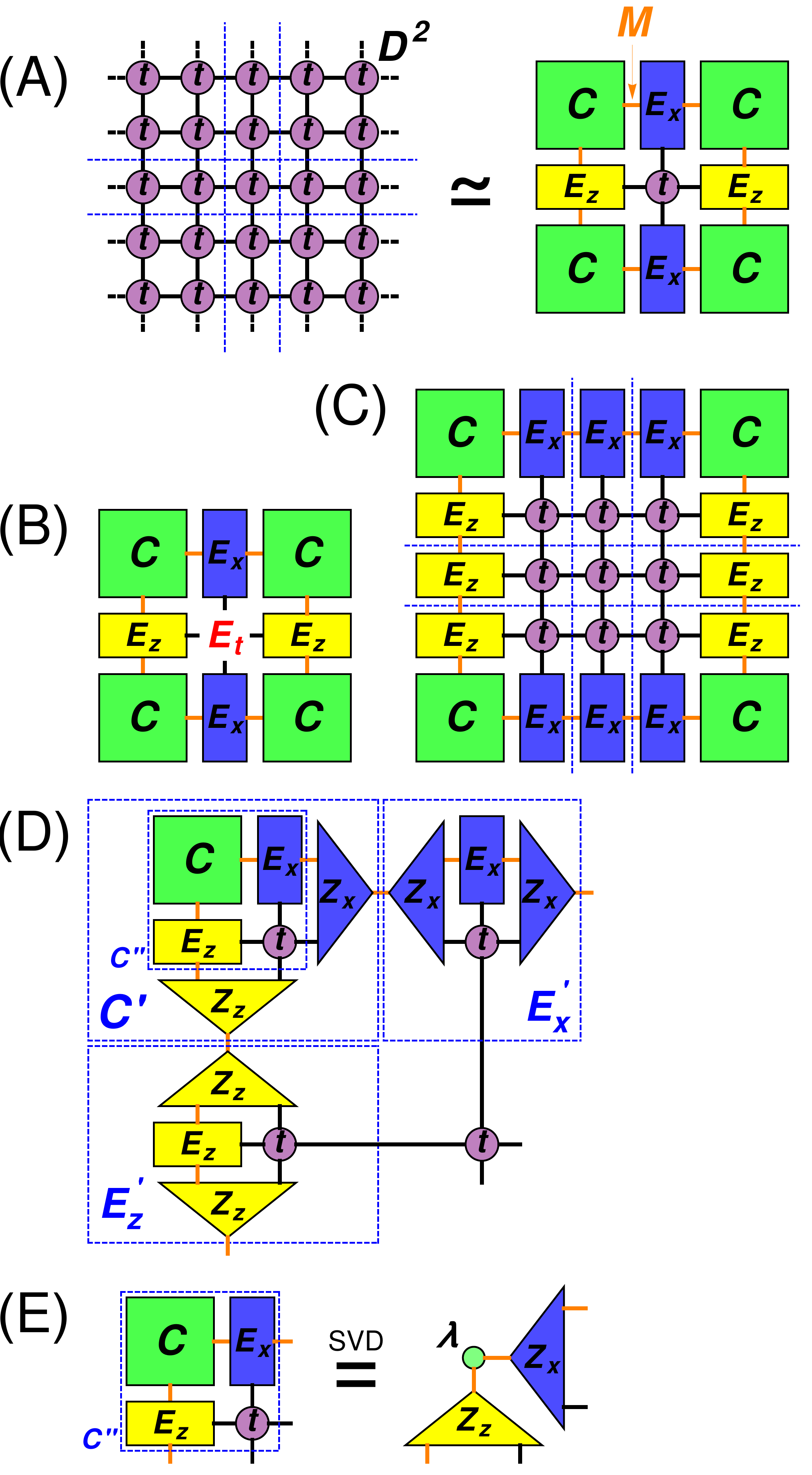}
\caption{
In A,
planar version of the partition function in Fig. \ref{fig:Tn}C.
From the point of view of the central tensor $t$,
this infinite network can be replaced by a finite effective one made of
a corner matrix $C$ and edge tensors $E_x$ and $E_z$. Each of them
represents its corresponding infinite sector delimited by the blue
dashed lines. The environmental tensors contract through bond indices
of dimension $M$.
In B,
the finite effective environment $E_t$ made of the finite environmental
tensors. With increasing $M$ it becomes the exact one in Fig.
\ref{fig:Envt}A.
In C,
a network equivalent to the networks in panel A.
Here the blue dashed lines separate enlarged environmental tensors.
In D,
the enlarged tensors are renormalized by isometries $Z_x$ and $Z_z$ to
new tensors $C'$, $E'_x$ and $E'_z$ back with the environmental bond
dimension $M$.
In E,
the isometries are obtained from a singular value decomposition (SVD)
of the enlarged corner: $C''=Z_z\lambda Z_x^\dag$.
}
\label{fig:CMR}
\end{figure}
%%%%%%%%%%%%%%%%%%%%%%%%%%%%%%%%%%%%%%%%%%%%%%%%%%%%%%%%%%%%%%%%%%%%%%%%%%%%

An infinite network, like the one in Fig. \ref{fig:Tn}C, cannot be
contracted exactly but, fortunately, what we often need is not this
number, but a tensor environment for a few sites of interest like,
for instance,
the environment $E_t$ in Fig. \ref{fig:Envt}A.
From the point of view of the removed $t$, its exact infinite
environment can be substituted with a finite effective one
made of finite corner matrices $C$ and edge tensors $E_x$ and $E_z$,
see Fig. \ref{fig:CMR}.
The environmental tensors $C$ and $E$ are contracted with each other by
environmental bond indices of dimension $M$. By increasing $M$
the effective $E_t$ can be converged towards the exact one in a
systematic way. When the correlation length is finite the convergence
is reached exponentially at a finite $M$. At a critical point,
even though the correlation length $\xi(M)$ remains finite for any
finite $M$, it quickly diverges with a power of $M$
making local observables and correlations up to the distance
$\xi(M)$ converge to their exact values \cite{self,var}.

The finite tensors $C$ and $E$ represent infinite sectors of the network
in Fig. \ref{fig:CMR}A. The tensors are converged by iterating
the corner matrix renormalization in Figs. \ref{fig:CMR}C-E.
In every renormalization step, the corner $C$ is enlarged to $C''$.
This operation represents the top-left corner sector in Fig.
\ref{fig:CMR}A absorbing one more layer of tensors $t$. The enlarged
$C''$ is subject to singular value decomposition
$C''=Z_z\lambda Z_x^\dag$, see Fig. \ref{fig:CMR}E.
$\lambda$ is truncated to $M$ largest singular values and the
unitaries $Z_z$ and $Z_x$ to the corresponding isometries. The
isometries renormalize $C''$ and the enlarged edge tensors to a new
corner $C'$ and edges $E'$, respectively.
The whole procedure is iterated until convergence.

The numerical cost of converging the environmental tensors is
${\cal O}\left[M^3(D^2)^3\right]$, where $D^2$ is the bond dimension
of $t$. The cost of calculating $E_{T_n}$ can be reduced to
${\cal O}\left(M^2D^6,M^3D^4\right)$ if one goes directly from
the environmental tensors to $E_{T_n}$ without calculating the
intermediate $E_t$.

%%%%%%%%%%%%%%%%%%%%%%%%%%%%%%%%%%%%%%%%%%%%%%%%%%%%%%%%%%%%%%%%%%%%%%%%%%
\section{Error estimate}
\label{sec:Error}
%%%%%%%%%%%%%%%%%%%%%%%%%%%%%%%%%%%%%%%%%%%%%%%%%%%%%%%%%%%%%%%%%%%%%%%%%%

Observables should be converged not only in $M$ but also in $D$.
A modest $D\simeq 7$ is sufficient in the 2D quantum Ising model
\cite{var}, but in realistic models rather than full convergence we
would expect to get close enough to it to make a reliable extrapolation
with $1/D\to0$. However, the raw $1/D$ may be not the most reliable
small parameter for the extrapolation \cite{CorbozHubbard}.
For instance, the PEPO \textit{Ansatz} may
not change much between $D$ and $D+1$ but then suddenly improve for
$D+2$ making the dependence of observables on $1/D$ rough.
A more direct measure of the actual error inflicted by a finite $D$
would make the dependence smoother and the extrapolation more reliable.

%%%%%%%%%%%%%%%%%%%%%%%%%%%%%%%%%%%%%%%%%%%%%%%%%%%%%%%%%%%%%%%%%%%%%%%%%%%%%%%%%%
\begin{figure}[t!]
\includegraphics[width=0.7\columnwidth,clip=true]{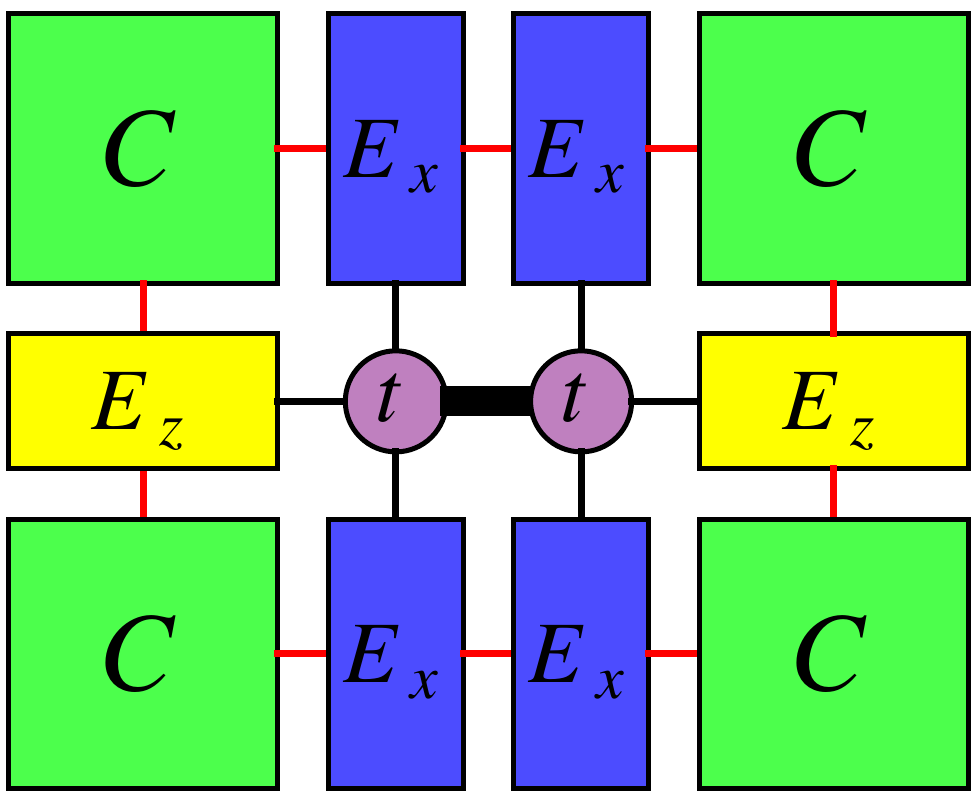}
\caption{
The network used to estimate the error along the central $a$-bond
inflicted by the isometries $W^x_m$ with the bond dimension $D$.
Its contraction is a number $N^x_D$.
When the bond dimension of the isometries on the central bond is
enlarged to $D'>D$ and the enlarged isometries on this bond are
optimized the number becomes $N^x_{D'}$.
For large enough $D'$ it converges to $N^x_\infty$.
The relative error $e_x$ is given by Eq. \eqref{ex}.
}
\label{fig:Error}
\end{figure}
%%%%%%%%%%%%%%%%%%%%%%%%%%%%%%%%%%%%%%%%%%%%%%%%%%%%%%%%%%%%%%%%%%%%%%%%%%%%

The measure can be constructed in a similar way as for the
zero-temperature PEPS \cite{CorbozHubbard}. Figure \ref{fig:Error}
shows the network used to estimate the error inflicted by the
isometries $W^x$ with the bond dimension $D$. This network is a number
$N^x_D$. When the bond dimension of the isometries on the central bond
is enlarged to $D'>D$ and the enlarged isometries on this bond are
optimized, then the number becomes $N^x_{D'}$.
It converges to $N^x_\infty$ for a large enough $D'$.
In our calculations $D'=4D$ proved to be sufficient.
The relative error is given by
\be
e_x(D)=(N^x_\infty-N^x_D)/N^x_\infty.
\label{ex}
\ee
In a similar way we obtain the error inflicted by isometries $W^z$ on
a bond along the $b$ axis:
\be
e_z(D)=(N^z_\infty-N^z_D)/N^z_\infty.
\ee

%%%%%%%%%%%%%%%%%%%%%%%%%%%%%%%%%%%%%%%%%%%%%%%%%%%%%%%%%%%%%%%%%%%%%%%%%%
\section{Figure of merit}
\label{sec:Px}
%%%%%%%%%%%%%%%%%%%%%%%%%%%%%%%%%%%%%%%%%%%%%%%%%%%%%%%%%%%%%%%%%%%%%%%%%%

The algorithm optimizes each isometry $W^x_m$ to maximize its overlap
with its environment $E_{W^x_m}$. As the overlap is proportional to
the partition function $Z$, the optimization aims at maximizing $Z$.
We will argue that in the compass model maximizing $Z$ is equivalent
to minimizing the error inflicted on $Z$ by the isometry.

Indeed,
the $n$ layers of isometries $W^x_1$,...,$W^x_n$ make a tree tensor
network like the one shown in Fig. \ref{fig:TTN} in case of $n=3$
layers. The whole TTN is also an isometry to be called $W_x$.
(We note in passing that in principle the TTN could be replaced with
a more general tensor network like the one in Ref. \cite{ramsetal},
but it is not clear at the time of writing how to perform its
variational optimization with full tensor environment.) In the PEPO
\textit{Ansatz} for the gate $U(\beta)$ in Fig. \ref{fig:Tm}B
on every bond along the $a$ axis there are two isometries $W_x$ that
combine into a projector $P_x=W_xW_x^\dag$. We want to minimize the
error inflicted on the partition function by $P_x$.

%%%%%%%%%%%%%%%%%%%%%%%%%%%%%%%%%%%%%%%%%%%%%%%%%%%%%%%%%%%%%%%%%%%%%%%%%%%%%%%
\begin{figure}[t!]
\includegraphics[width=0.99\columnwidth,clip=true]{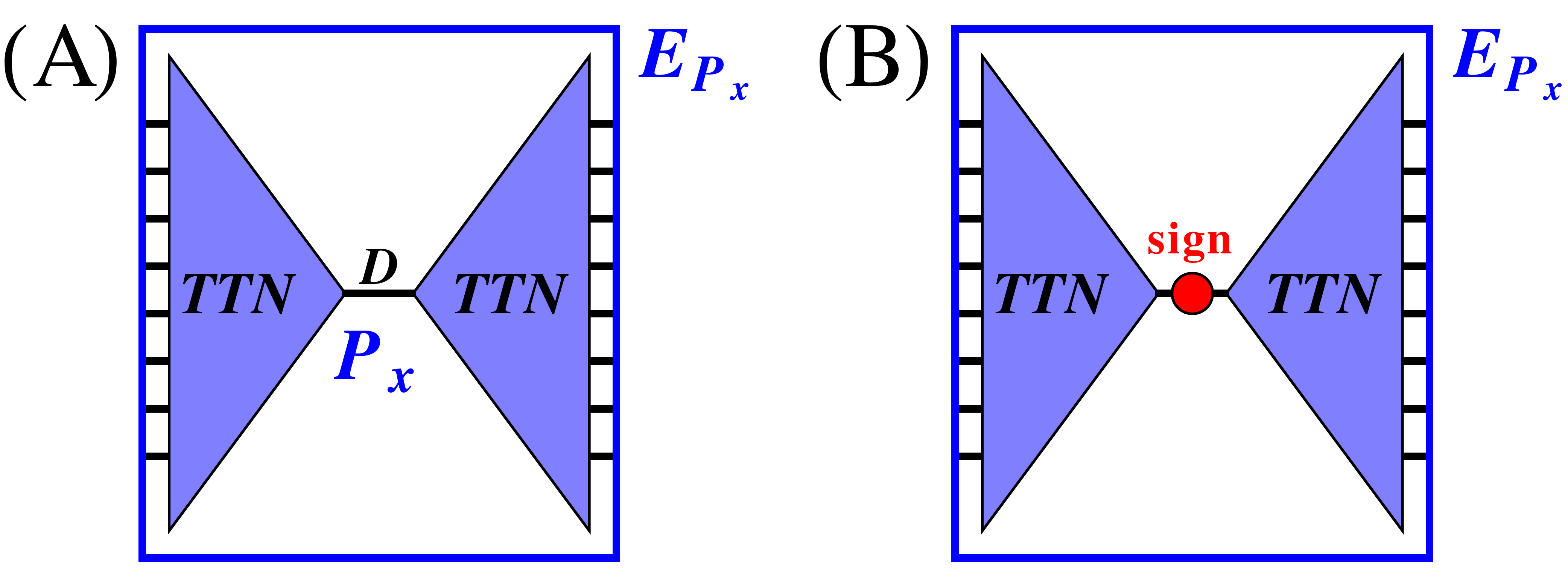}
\caption{
In A,
$n$ layers of isometries $W^x_1$,...,$W^x_n$ make a TTN,
see Fig. \ref{fig:TTN} in case of $n=3$ layers.
The TTN is also an isometry to be called $W_x$.
Two TTNs make a projector $P_x=W_xW^\dag_x$.
In B,
the sign matrix (\ref{S}) is inserted into the central bond.
After this insertion the isometries $W^x_m$ can be updated by maximizing
their overlaps with their respective environments even when
$E_{P_x}$ is not positive-semidefinite.
}
\label{fig:Px}
\end{figure}
%%%%%%%%%%%%%%%%%%%%%%%%%%%%%%%%%%%%%%%%%%%%%%%%%%%%%%%%%%%%%%%%%%%%%%%%%%%%

The partition function in Fig. \ref{fig:Tn} can be represented by the
effective network in Fig. \ref{fig:Error}.
We focus on the two projectors $P_x$ on the central bond,
one in each of the two layers $U(\beta)$.
Given the left-right symmetry of this network reflecting the symmetry
of the compass model, it can be shown that
$E_{P_x}={\cal M}{\cal M}^\dag$, where ${\cal M}$ is a huge matrix
representing the left half of Fig. \ref{fig:Error}. The environment is
symmetric and positive-semidefinite, hence the partition function is
distorted least by the projector $P_x$ that maximizes its contraction
with $E_{P_x}$, see Fig. \ref{fig:Px}.
This optimal projector is made of isometries that in turn maximize
their overlaps with the respective environments.

In order to put this simple result in a more general context,
let us consider now $E_{P_x}$ that is not positive-semidefinite but
is still symmetric. Since the $P_x$ to be contracted with $E_{P_x}$ is
symmetric, hence only the symmetric part of $E_{P_x}$ matters anyway.
Now the least distortive projector is no longer the one on the largest
eigenvalues of $E_{P_x}$, but that on the eigenvalues with the largest
magnitudes. The huge $E_{P_x}$ can be neither diagonalized nor even
calculated, but the $D\times D$ matrix $e_{P_x}$ obtained after cutting
the central $D$-bond in Fig. \ref{fig:Px} is $E_{P_x}$ projected on a
$D$-dimensional subspace.
This matrix can be efficiently calculated and diagonalized,
\be
e_{P_x}=\sum_{\alpha=1}^D|\alpha\rangle \lambda_\alpha \langle\alpha|,
\ee
and we can construct its sign operator,
\be
{\rm sign}\left(e_{P_x}\right) = \sum_{\alpha=1}^D |\alpha\rangle\,
{\rm sign}\left(\lambda_\alpha\right) \langle\alpha|.
\label{S}
\ee
Inserting this sign into the cut $D$-bond in Fig. \ref{fig:Px} is
equivalent to replacing the eigenvalues of $E_{P_x}$ by their magnitudes.
With the inserted sign,
the least distortive isometries $W^x_m$ are again those that maximize
their overlaps with their respective environments.
The sign insertion is a redundant null operation when $E_{P_x}\geq0$,
like in the quantum compass or quantum Ising models,
but it proves essential in the fermionic Hubbard model~\cite{Cza16}.

%%%%%%%%%%%%%%%%%%%%%%%%%%%%%%%%%%%%%%%%%%%%%%%%%%%%%%%%%%%%%%%%%%%%%%%%%%%%

\end{document}